\documentclass{article}
\usepackage[utf8]{inputenc}

\pdfoutput=1
\usepackage{amssymb}
\usepackage{amsmath}
\usepackage[dvips]{graphicx}
\usepackage{setspace}
\usepackage{slashed}
\usepackage{mathtools}
\usepackage{amsfonts}
\usepackage{fancyhdr}
\usepackage{xcolor}
\usepackage{graphicx}
\usepackage{rotating}
\usepackage{comment}
\usepackage{color}
\usepackage{subcaption}
\usepackage[percent]{overpic}
\usepackage{cite}
\usepackage{braket}
\usepackage{moresize}
\usepackage{dsfont}

\definecolor{darkgreen}{rgb}{0,0.5,0}
\definecolor{darkblue}{rgb}{0,0,0.6}
\definecolor{purple}{rgb}{0.4,.2,0.7}
\newcommand{\p}{\partial}

\newcommand{\be}{\begin{equation}}
\newcommand{\ee}{\end{equation}}

\usepackage[colorlinks=true,citecolor=darkgreen,linkcolor=black,urlcolor=purple]{hyperref}

\usepackage{pdfsync}
\makeatletter
\newcommand*{\defeq}{\mathrel{\rlap{%
                     \raisebox{0.3ex}{$\m@th\cdot$}}%
                     \raisebox{-0.3ex}{$\m@th\cdot$}}%
                     =} 
\makeatother

\def\be{\begin{eqnarray}}
\def\ee{\end{eqnarray}}

\newcommand{\bea}{\begin{eqnarray}}
\newcommand{\eea}{\end{eqnarray}}
\def\ben{\begin{equation}}
\def\een{\end{equation}}

    \let\p=\phi \let\r=v

\def\be{\begin{equation}}
\def\ee{\end{equation}}
\def\ba{\begin{eqnarray}}
\def\ea{\end{eqnarray}}

\def\bal#1\eal{\begin{align}#1\end{align}}
\def\bs#1\es{\begin{split}#1\end{split}}

\renewcommand{\p}{\partial}

\allowdisplaybreaks  % allow page breaks in displayed eqs

\numberwithin{equation}{section}
\def\p{{\phi}}

\def\be{\begin{equation}}
\def\ee{\end{equation}}
\def\ba{\begin{eqnarray}}
\def\ea{\end{eqnarray}}
\def\bal#1\eal{\begin{align}#1\end{align}}

\def\r{\rightarrow}

\def\r{\right}

\usepackage{tikz}
\usetikzlibrary{positioning,arrows}
\usetikzlibrary{decorations.pathmorphing}
\usetikzlibrary{decorations.markings}
\tikzset{
particle/.style={postaction={decorate}},
graviton/.style={decorate, decoration={snake, amplitude=0.8 mm, segment length=1.5 mm, pre length=0.8 mm, post length=0.8 mm}},
photon/.style={
        decoration={complete sines, amplitude=0.15cm, segment length=0.2cm},
        decorate    
    },
gluon/.style={
        decoration={coil, aspect=0.75, mirror, segment length=1.5mm},
        decorate
    }
}
 
\def \be {\begin{equation}}
\def \ee {\end{equation}}

\renewcommand{\p}{\partial}

\usepackage{framed}

\begin{document}
\onehalfspacing

\begin{center}

~
\vskip5mm

{\LARGE  {
Euclidean wormholes in holographic RG flows
}}

\vskip7mm
Jeevan Chandra

\vskip5mm

{\it Department of Physics, Cornell University, Ithaca, New York, USA 
}

\vskip5mm

\end{center}

\vspace{2mm}

\begin{abstract}
\noindent

We describe a one-parameter family of Euclidean wormhole solutions with the topology of a compact hyperbolic space times an interval in Einstein gravity minimally coupled to a massless scalar field in AdS$_{d+1}$ commonly referred to as Einstein-dilaton gravity. These solutions are locally described by the same metric and dilaton profile as the single-boundary Janus domain wall solutions in the same theory which are usually studied in the context of holographic RG flows. The wormholes compute the averaged product of partition functions of CFTs on either boundary deformed by different marginal couplings to the scalar operator dual to the dilaton. We observe that the renormalised volumes of these wormholes increase monotonically with the difference in the marginal couplings on the boundary thereby showing that the pair of CFTs on the boundaries get increasingly decorrelated as the difference in the marginal couplings increases. We use the partition functions of the three-dimensional wormhole solutions to determine the variance of the OPE data of local operators between the marginally deformed 2d CFTs and quantify how the variance decays with the difference in marginal couplings. In addition, a family of wormholes sourced by a thin shell of dust determine how the variance of the matrix elements of the dual line defect decays with the difference in marginal couplings. Applying the GKPW dictionary to wormholes, we compute averages of integrated dilaton correlators treating the wormhole amplitude as a functional of the dilaton sources. We observe that the crossed two-point correlators with a dilaton insertion on either boundary decay montonically with the difference in marginal couplings consistent with the observation that the CFTs increasingly decorrelate as the difference in marginal couplings grows.

 \end{abstract}

\pagebreak
\pagestyle{plain}

\setcounter{tocdepth}{2}
{}
\vfill
\tableofcontents

%\setcounter{tocdepth}{1}
%\newpage

\date{}
%\maketitle

\newcommand{\btau}{\bar{\tau}}

\section{Introduction}

Euclidean wormholes are connected contributions to the gravitational path integral joining disconnected boundary components \cite{Maldacena:2004rf,Witten:1999xp}.
In recent years, there has been a lot of progress in the study of such wormhole solutions in three-dimensional gravity and in understanding their boundary interpretation. Off-shell wormholes, in particular those with the topology of torus times interval have been shown to contain information about the energy level spacing statistics of 2d CFT \cite{Cotler:2020ugk,DiUbaldo:2023qli}. On the other hand, on-shell wormholes which solve Einstein's equations in vacuum or in the presence of matter worldlines have been shown to encode statistics of OPE data of holographic 2d CFTs \cite{Belin:2020hea, Chandra:2022bqq} thereby realizing that 3d gravity obeys a version of the Eigenstate Thermalization Hypothesis (ETH) consistent with Virasoro symmetry \cite{Collier:2019weq}. The consistency with ETH was verified at the classical level and 1-loop order in \cite{Chandra:2022bqq} and to all loop orders using Virasoro TQFT in \cite{Collier_2023,Collier:2024mgv} (see also \cite{Abajian:2023bqv,Chandra:2023dgq,Chandra:2023rhx,Yan:2023rjh,deBoer:2024kat} for related works). Recently, there has been a proposal to reproduce both the on-shell and off-shell wormhole amplitudes using a matrix-tensor model whose potential is determined by the bootstrap constraints \cite{Belin:2023efa,Jafferis:2024jkb}. This idea generalizes the result of Saad, Shenker and Stanford \cite{Saad:2019lba} who showed that wormhole amplitudes in JT gravity match with observables obtained by averaging over ensembles of random matrix Hamiltonians, to three dimensions by incorporating bootstrap constraints into the matrix integrals.  

A similar story about the relation between wormholes and statistics is expected to generalize to higher dimensions: Statistics of black hole microstates are reproduced by wormhole amplitudes. It is however hard to calculate wormhole amplitudes in higher dimensions at the same level of detail as in two and three dimensions. A tractable model in higher dimensions seems to be provided by the black hole and wormhole solutions sourced by spherically symmetric thin shells of dust particles in which case the wormhole amplitudes have been shown to approximately reproduce the Gaussian statistics of the corresponding black hole microstate data \cite{Chandra:2022fwi,Sasieta:2022ksu,Balasubramanian:2022gmo,Bah_2023,Chandra:2024vhm}. This is not surprising since the requirement of spherical symmetry effectively reduces the higher dimensional problem to a two-dimensional problem. See \cite{Maldacena:2004rf,Cotler_2021,Marolf:2021kjc} for examples of higher dimensional non-spherically symmetric wormholes in pure gravity or gravity coupled to matter fields where the stability of wormholes to field theoretic negative modes and brane nucleation has been studied.

In this paper, we describe wormhole solutions in models used to describe holographic RG flows. See also \cite{Ghodsi:2022umc,Betzios:2019rds} for a general discussion and other perspectives on the relationship between Euclidean wormholes and RG flows between QFTs placed on hyperbolic spaces. The general goal of the present work is to construct wormholes which correlate observables in different CFTs and to quantify how such wormholes which interpolate between different boundary theories affect the statistics of CFT data. To this end, we work with perhaps the simplest model of a holographic RG flow: Einstein gravity in AdS$_{d+1}$ minimally coupled to a massless scalar field commonly referred to as Einstein-dilaton gravity. This model famously admits single-boundary domain wall solutions which for $d=2$ and $d=4$ can be obtained by dimensionally reducing the Janus domain wall solutions of IIB supergravity \cite{Bak:2003jk,Bak:2007jm}. However, following the convention of \cite{Papadimitriou:2004rz}, we refer to such domain wall solutions in any dimension as Janus solutions. The Janus solutions are dual to a conformal interface on the boundary across which the coupling to the marginal scalar dual to the dilaton jumps abruptly. In the famous example of the duality between $\mathcal{N}=4$ SYM and IIB supergravity on AdS$_5 \times S^5$ \cite{Maldacena:1997re}, the dilaton in the gravitational theory governs the Yang-Mills coupling on the boundary. In this example, the Janus solutions of AdS$_5$ are dual to conformal interfaces in $\mathcal{N}=4$ SYM across which the Yang-Mills coupling jumps abruptly.

We construct two-boundary wormholes with the topology of a compact hyperbolic surface times an interval which are locally described by the single-boundary Janus solutions. They calculate averages of observables in two copies of CFT deformed by different marginal couplings. By computing the wormhole amplitudes, we show that the correlation between the partition functions of the two copies decreases monotonically with the difference in marginal couplings in the $G_N\to 0$ limit,
\begin{equation} \label{defvarintro}
  \overline{Z[\Sigma_d]^{\phi_+} Z[\Sigma_d]^{\phi_-}}=\exp\left(- \frac{d}{8\pi G_N}F(b)V(\Sigma_d)\right) \overline{Z[\Sigma_d] Z[\Sigma_d]}
\end{equation}
where $\Sigma_d$ is a $d$-dimensional compact hyperbolic surface with finite volume $V(\Sigma_d)$, $\phi_\pm$ are the marginal couplings on either boundary. $b$ parametrises the difference in these couplings and $F(b)$ is the difference in the renormalised volumes of the deformed and undeformed wormholes whose closed-form expression is given in \eqref{voleven} and \eqref{volodd} increases monotonically with $b$ and also shown to increase monotonically with the difference $\Delta \phi \equiv |\phi_+-\phi_-|$. The average $\overline{Z[\Sigma_d] Z[\Sigma_d]}$ is the variance when there is no difference in marginal coupling. Thus, the result \eqref{defvarintro} illustrates that the correlation decays with the difference in marginal couplings. Note that when the dual is a large-$N$ gauge theory, $G_N \sim \frac{1}{N^2}$ and if we assume that the difference in couplings is small, the result from the wormhole amplitude \eqref{defvarintro} shows that the correlation decays as 
\begin{equation}
  \frac{ \overline{Z[\Sigma_d]^{\phi_+} Z[\Sigma_d]^{\phi_-}}}{\overline{Z[\Sigma_d] Z[\Sigma_d]}}\approx \exp\left (-\# (\Delta \phi)^2N^2 V(\Sigma_d)\right )  
\end{equation}
where $\#$ is a proportionality factor that only depends on $d$. For example, in AdS$_5$, it is $\frac{3}{32\pi^2}$ as shown in \eqref{corrsmall}.
In three dimensions, the result \eqref{defvarintro} takes a simpler form,
\begin{equation}  \label{3dcorrintro}
  \overline{Z[\Sigma_{g,n}]^{\phi_0+\frac{\Delta \phi}{\sqrt{2}}}Z[\Sigma_{g,n}]^{\phi_0-\frac{\Delta \phi}{\sqrt{2}}}}=(\cosh(\Delta \phi))^{\frac{c}{6}\chi(\Sigma_{g,n})}\overline{Z[\Sigma_{g,n}]Z[\Sigma_{g,n}]}
\end{equation}
where $\Sigma_{g,n}$ is a compact hyperbolic Riemann surface with genus $g$ and $n$ cone points and $\chi(\Sigma_{g,n})$ is its Euler characteristic which is negative. $c=\frac{3}{2G_N}$ is the central charge of either CFT. This result applied to $\Sigma_{g,n}=\Sigma_{0,3}$ immediately gives the expression for the deformed variance of the OPE data of heavy local operators below the black hole threshold at large $c$,
\begin{align}
   Z_{\text{grav}}\left [\qquad  \vcenter{\hbox{
 \begin{overpic}[scale=0.6,grid=false]{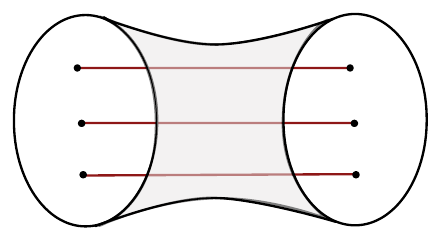}
 \put (-15,25) {$\phi_0+\frac{\Delta \phi}{\sqrt{2}}$}
 \put (50,40) {$i$}
 \put (50,28) {$j$}
 \put (50,15) {$k$}
 \put (90,25) {$\phi_0-\frac{\Delta \phi}{\sqrt{2}}$}
    \end{overpic}
 }} \qquad \right ]=& |C_0(h_i,h_j,h_k)|^2 \left(\cosh(\Delta \phi) \right )^{-\frac{c}{3}(\eta_i+\eta_j+\eta_k-1)}  \\ \notag &= \overline{c_{ijk}^{\phi_0+\frac{\Delta \phi}{\sqrt{2}}}c_{ijk}^{\phi_0-\frac{\Delta \phi}{\sqrt{2}}}}
\end{align}
where $h=\frac{c}{6}\eta(1-\eta)$ is the conformal weight of the operator and $\eta \in (0,\frac{1}{2})$ parametrises the conical deficit angle. $C_0$ is the smooth function of conformal weights determining the variance of the OPE coefficients in the undeformed CFT whose explicit form can be found for instance in \cite{Collier:2019weq} but is not important for the present purpose. The general expression for the deformed variance expected to hold even for operators above the black hole threshold is given in \eqref{OPElocal}. We refer the reader to the discussion below \eqref{OPElocal} for comments on how the dilatonic deformation is insensitive to weights of operators above the black hole threshold. Generalizing \eqref{3dcorrintro} to Riemann surfaces backreacted on by line defects, we compute the deformed variance of the matrix elements of the line defect $D_\Sigma$ in \eqref{linedef} dual to a dust shell in 3d gravity \cite{Chandra:2024vhm} at large $c$,
\begin{align}
 Z_{\text{grav}}\left [\quad \qquad
\vcenter{\hbox{
\begin{overpic}[width=1.8in,grid=false]{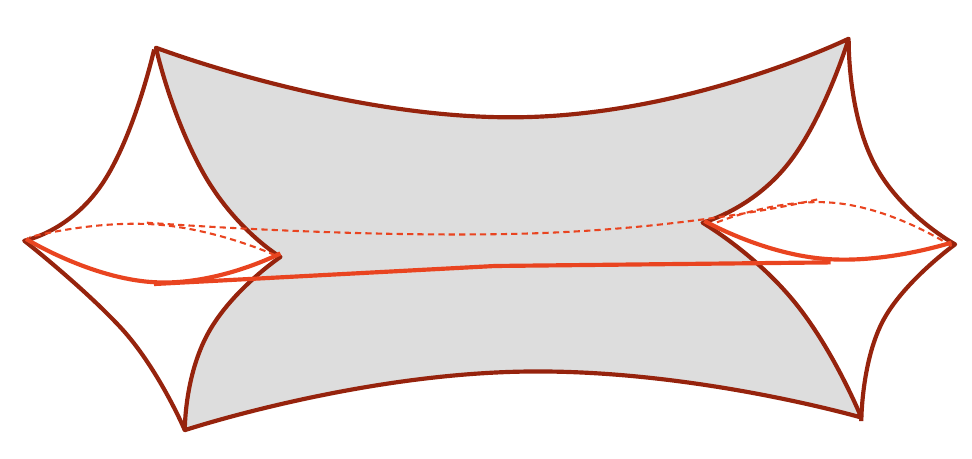}
 \put (-25,20) {$\phi_0+\frac{\Delta \phi}{\sqrt{2}}$}
\put (50,3) {$j$}
\put (50,38) {$i$}
\put (12,25) {$D_\Sigma$}
\put (83,25) {$D_\Sigma$}
 \put (100,20) {$\phi_0-\frac{\Delta \phi}{\sqrt{2}}$}
\end{overpic}
}}\quad \qquad \right ] &=|C_0^D(h_i,h_j;m)|^2( \cosh(\Delta \phi))^{-\frac{c}{3}(\eta_i+\eta_j+\frac{m}{2}-1)}\\ \notag & = \overline{\bra{i}D_\Sigma \ket{j}^{\phi_0+\frac{\Delta \phi}{\sqrt{2}}}\bra{j}D_\Sigma^\dagger \ket{i}^{\phi_0-\frac{\Delta \phi}{\sqrt{2}}}}
\end{align}
where $m$ is the mass parameter of the shell. The explicit form of the function $C_0^D$ which determines the variance in the undeformed CFT can be found in \cite{Chandra:2024vhm} but is unimportant for this work. We refer the reader to Appendix \ref{thinshellapp} for a quick review of the relevant Liouville formulae that goes into deriving this result. Finally, treating the wormhole amplitude as the generating functional for correlation functions of the marginal scalar dual to the dilaton, we compute averages of integrated dilaton correlators and understand their features in sections \ref{dil3d} and \ref{dilhigherd}. 

We would like to emphasize that the formal averages over two-copy observables in this paper like the ones in the above paragraph could be interpreted as averages over small windows around the marginal couplings on the two copies. This is in addition to the microcanonical averaging for the OPE data necessary to make sense of these expressions. In fact, by including a marginal scalar in the spectrum, the formal ensemble of OPE data of 2d CFTs proposed in \cite{Chandra:2022bqq} to reproduce multi-boundary wormhole amplitudes of 3d gravity could be realised as an average over the marginal coupling to the scalar. Note that the wormhole solutions to 3d Einstein gravity are also solutions to 3d Einstein-dilaton gravity with a uniform dilaton profile. So, the ensemble described by averaging over marginal couplings in a small window around the same value on all the copies reproduces these wormhole amplitudes. 

The paper is organised as follows: In section \ref{3dwormholes}, we describe three-dimensional wormhole solutions constructed from Janus domain wall solutions in 3d gravity, compute the wormhole amplitudes and understand the consequences on statistics of 2d CFT data. In section \ref{higherdwormholes}, we generalize the analysis to higher dimensions by constructing wormhole solutions using Janus domain walls in AdS$_{d+1}$. Finally, in section \ref{discussion}, we conclude and discuss some interesting future directions. There are a couple of appendices with additional details.

\section{Wormholes in three-dimensional holographic RG flows} \label{3dwormholes}

We start by reviewing the three-dimensional Janus solutions. We then turn these solutions into a one-parameter family of two-boundary wormholes with the topology of $\Sigma_{g,n}\times \mathbb{R}$ which are deformations of the Maldacena-Maoz wormholes \cite{Maldacena:2004rf} of the same topology. We compute the volumes of these wormholes and show that they increase montonically under the deformation at a rate governed by the magnitude of the Euler characteristic $\chi(\Sigma_{g,n})$ of the Riemann surface on the boundary. We then show that the wormhole amplitudes between flat or hyperbolic boundary metrics are governed by these volumes. We interpret the wormhole amplitudes in terms of averages of observables between the two CFTs on either boundary which are both marginally deformed by the scalar dual to the dilaton but by different marginal couplings. We then use this observation to propose how the statistics of 2d CFT data is modified due to the deformation. Finally, we use the wormhole amplitudes to compute integrated dilaton correlators in the background of these wormholes.

\subsection{Review: Janus domain walls in three dimensions}

The Janus domain wall solutions are single-boundary solutions to Einstein's equations minimally coupled to a massless scalar field which we shall refer to as the dilaton. These solutions have been studied extensively in the literature and were constructed in \cite{Bak:2007jm} by dimensionally reducing $10d$ domain wall solutions of IIB supergravity on AdS$_3 \times S^3 \times M_4$ where $M_4$ is a compact 4-manifold usually chosen to be $K3$ or $T^4$. We review some of the important features of these solutions. In Euclidean signature, the action is given by
\begin{equation} \label{action3d}
    I=-\frac{1}{16\pi G_N}\int \sqrt{g}(R+2-(\nabla \phi)^2)-\frac{1}{8\pi G_N}\int\sqrt{h}(K-1)
\end{equation}
where we have set $\ell_{\text{AdS}}=1$ and the cosmological constant $\Lambda=-1$.
Einstein's equations take the form
\begin{equation}
    R_{\mu \nu}-\frac{1}{2}g_{\mu \nu}R-g_{\mu \nu}=\partial_\mu \phi \partial_\nu \phi -\frac{1}{2} g_{\mu \nu} (\nabla \phi)^2
\end{equation}
The RHS is the stress tensor for the massless scalar field. Taking the trace, we see that 
\begin{equation}
    R=-6+(\nabla \phi)^2
\end{equation}
using which the above Einstein's equations can be rewritten as
\begin{equation}
    R_{\mu \nu}+2g_{\mu \nu}=\partial_\mu \phi \partial_\nu \phi
\end{equation}
The Janus solution for the metric can be expressed as
\begin{equation}
    ds^2=d\rho^2+f(\rho)d\Sigma^2
\end{equation}
where $d\Sigma^2=\frac{dx^2+dy^2}{y^2}$ is the Poincare metric on the upper half plane and the warp factor $f(\rho)$ is given by
\begin{equation} \label{warp3d}
    f(\rho)=\frac{1}{2}(1+a^2 \cosh(2\rho))
\end{equation}
and the parameter $a \in (0,1]$. When $a=1$, we get the familiar warp factor $f(\rho)=\cosh^2(\rho)$ corresponding to hyperbolic slicing of AdS$_3$. Note that the warp factor \eqref{warp3d} implies that the local curvature radius of the Janus solution changes smoothly along the radial direction. In this sense, the Janus solution describes a holographic RG flow. However, note that because of the $\mathbb{Z}_2$ symmetry of the metric under $\rho \to -\rho$, it means that the asymptotic curvature radius is the same for both the boundaries. In the dual description, this means the central charge of the CFT does not jump across the interface. It is important to note that such an RG flow solution with the warp factor depending only on the radial coordinate breaks the isometries of AdS$_3$ retaining however the isometries of the transverse AdS$_2$.

Now, we turn to the profile of the dilaton. The dilaton satisfies the massless Klein-Gordon equation,
\begin{equation}
    \Box \phi= \frac{1}{\sqrt{g}}(\sqrt{g}g^{\mu \nu}\partial_\mu \phi)_{,\nu}=0
\end{equation}
The profile for the dilaton using the ansatz that it depends only on the $\rho$ coordinate is given by
\begin{equation}
    \phi(\rho) =\phi_0+\sqrt{2}\tanh^{-1}\left (\sqrt{\frac{1-a^2}{1+a^2}}\tanh(\rho) \right )
\end{equation}
The profile interpolates between the asymptotic values 
\begin{equation}
    \phi_\pm =\phi_0\pm \sqrt{2}\tanh^{-1}\left (\sqrt{\frac{1-a^2}{1+a^2}} \right )
\end{equation}
The presence of the undetermined constant $\phi_0$ in the dilaton profile is expected since the action has a shift symmetry $\phi \to \phi + \text{const}$.
This solution is dual to a conformal interface in the dual CFT which in these coordinates is at $y=0$ on the boundary separating the two half-planes. For later use, it is convenient to express the parameter $a$ in terms of the difference in the asymptotic values of the dilaton as
\begin{equation}
    \phi_\pm =\phi_0 \pm \frac{\Delta \phi}{\sqrt{2}}\implies a^2=\frac{1}{\cosh(\Delta \phi)}
\end{equation}

\subsection{Monotonicity of wormhole actions and volumes}

Now, we describe two-boundary Euclidean wormhole solutions with the topology of a compact smooth hyperbolic Riemann surface $\Sigma$ times an interval. They are locally described by the same metric and have the same dilaton profile as the single-boundary Janus solutions reviewed in the previous section. They are constructed by taking quotients of the single-boundary Janus solutions by a discrete subgroup of the PSL($2,\mathbb{R})$ isometry group of the transverse $\mathbb{H}_2$ which as we noted earlier is preserved by the Janus deformation. The discrete subgroup $\Gamma$ is referred to as a Fuchsian group and is chosen such that $\mathbb{H}_2/\Gamma \cong \Sigma$. Such a quotient does not affect the dilaton profile since it is uniform in the transverse directions.  As a result of the quotient by a Fuchsian group, these wormholes connect Riemann surfaces of the same geometric moduli\footnote{When the dilaton profile is uniform, we can construct two-boundary wormholes with different boundary moduli and more general multi-boundary wormholes by taking quotients of the single-boundary solutions which are locally $\mathbb{H}_3$ by discrete subgroups of their PSL$(2,\mathbb{C})$ isometry group called Kleinian groups. However, once the Janus deformation or more generally any RG flow is turned on where the dilaton has a non-trivial radial profile, the PSL$(2,\mathbb{C})$ isometry group is broken down to the PSL$(2,\mathbb{R}$) isometry group of the radial slices. This means that two-boundary wormholes with different moduli and multi-boundary wormholes cannot be constructed by taking quotients of the single-boundary solutions since the Kleinian groups used to construct them are not subgroups of the residual PSL($2,\mathbb{R})$ isometry group.}. 
 This one-parameter family of wormhole solutions contain the Maldacena-Maoz wormholes \cite{Maldacena:2004rf} of the corresponding topology at a special value of the parameter. So, we refer to them as deformations of the Maldacena-Maoz wormholes.

The metric on these wormholes in hyperbolic slicing coordinates takes the form,
\begin{equation}
  ds^2=d\rho^2+\frac{1}{2}(1+a^2 \cosh(2\rho))\gamma_{ij}dx^idx^j
\end{equation}
where $\gamma_{ij}dx^idx^j$ is the hyperbolic metric on the compact surface. Notice that $a=1$ corresponds to the metric on the Maldacena-Maoz wormholes. We now compute the wormhole amplitudes between hyperbolic boundary metrics on the compact Riemann surfaces\footnote{See for instance \cite{Miyaji:2015woj} for a similar calculation of the on-shell action but for the single-boundary Janus solutions and its relation to the quantum information metric which calculates the fidelity between quantum states.}.
To compute the regulated on-shell action, we fix the cutoffs at $\rho=\pm \rho_c$ where
\begin{equation}
  \rho_c=\log(\frac{2}{a\epsilon})
\end{equation}
so that the induced metrics on the boundaries are given to leading order in the cutoff parameter $\epsilon$ by the hyperbolic metrics $\frac{1}{\epsilon^2}\gamma_{ij}dx^idx^j$. The bulk term evaluates to give
\begin{equation}
    -\frac{1}{16\pi G_N}\int \sqrt{g}(R+2-(\nabla \phi)^2)=\frac{1}{4\pi G_N}\int_\Sigma \sqrt{\gamma} \left(\frac{1}{\epsilon^2}+ \log(\frac{2}{a\epsilon}) \right )
\end{equation}
The induced metric on the cutoff surfaces is
\begin{equation}
    ds^2_{\text{bdry}}=\left(\frac{1}{\epsilon^2}+\frac{1}{2}\right )\gamma_{ij}dx^i dx^j
\end{equation}
The trace of the extrinsic curvature tensor on constant-$\rho$ slices is given by
\begin{equation}
  K=\frac{2a^2 \sinh(2\rho)}{1+a^2\cosh(2\rho)}
\end{equation}
using which we see that the extrinsic curvature of the cutoff slices is independent of $a$ upto $O(\epsilon^2)$,
\begin{equation}
    K=2-\epsilon^2+O(\epsilon^3)
\end{equation}
The boundary term evaluates to give
\begin{equation}
    -\frac{1}{8\pi G_N}\int \sqrt{h}(K-1)=-\frac{1}{4\pi G_N}\int_\Sigma \sqrt{\gamma} \left (\frac{1}{\epsilon^2}-\frac{1}{2}\right )
\end{equation}
Thus, the regulated on-shell action for the wormhole reads
\begin{equation}
    I(a)=\frac{c}{6}\chi(-1+2\log (\frac{\epsilon}{2}))+\frac{c}{3}\chi \log(a)
\end{equation}
where we used the fact that the central charge $c$ if the dual CFT is given by $c=\frac{3}{2G_N}$. The second term in the action is the interesting term since it depends on $a$. Here, $\chi$ is the Euler characteristic of the smooth Riemann surface determined by its genus $g$ as $\chi=2-2g$ and related to its hyperbolic area by
\begin{equation}
  \int _\Sigma \sqrt{\gamma} = -2\pi \chi
\end{equation}
The Brown-York stress tensor \cite{Balasubramanian:1999re} on either boundary evaluates to give
\begin{equation}
  T_{ij}=\frac{c}{48\pi}R[\gamma]\gamma_{ij}
\end{equation}
where $R[\gamma]$ is the Ricci scalar for the hyperbolic metric $\gamma$ and has been scaled to $R[\gamma]=-2$ with the chosen ansatz for the 3d metric. The form of the stress tensor is consistent with the Weyl anomaly for 2d CFTs placed on curved surfaces. Importantly, we see that the stress tensor is independent of $a$ which means that the Janus-like deformation for any values of the couplings $\phi_{\pm}$ is marginal for the CFTs on either boundary at large $c$.

Notice that since the cutoff-dependent piece of the action is independent of $a$, the difference between the regulated actions of the deformed and undeformed wormholes is finite. We shall refer to this difference as the wormhole action $I_{\text{WH}}(a)$,
\begin{equation}
  I_{\text{WH}}(a)\equiv I(a)-I(1)=\frac{c}{3}\chi \log(a)
\end{equation}
Expressed in terms of the difference in the asymptotic values of the dilaton,
\begin{equation}
  I_{\text{WH}}(\Delta \phi) =-\frac{c}{6}\chi \log (\cosh(\Delta \phi))
\end{equation}
so we see that the wormhole action is non-negative and increases monotonically with $\Delta \phi$ at a rate controlled by the magnitude of the Euler characteristic since hyperbolic surfaces have $\chi<0$. This means the wormhole partition function defined as 
\begin{equation}
 Z_{\text{WH}}\equiv e^{-I_{\text{WH}}}
\end{equation}
decreases monotonically with the $\Delta \phi$. Note that by definition, $Z_{\text{WH}}$ is rescaled using the partition function of the undeformed wormhole. 
 Since the boundary terms evaluated on the cutoff surfaces are independent of $a$, the wormhole action matches with the difference in regulated volumes of the wormholes,
\begin{equation}
  I_{\text{WH}}(a) = \frac{1}{4\pi G}(V(a)-V(1)) \qquad V(a) \equiv \int _{\text{reg}}\sqrt{g}
\end{equation}
Inspired by this observation, we define the volumes of the 1-parameter family of wormhole solutions by the difference in the regulated volumes of the deformed and undeformed wormholes,
\begin{equation}
 V_{\text{WH}}(a)\equiv V(a)-V(1)=2\pi \chi \log(a)
\end{equation}
Since $I_{\text{WH}}(a)=\frac{V_{\text{WH}}(a)}{4\pi G}$, the renormalised volumes enjoy the same positivity and monotonicity properties as the wormhole actions or partition functions. We can express the wormhole actions and volumes as solutions to the differential equation,
\begin{equation}
  a\frac{\partial I_{\text{WH}}}{\partial a}=\frac{c}{3}\chi \qquad a\frac{\partial V_{\text{WH}}}{\partial a}=2\pi \chi
\end{equation}
governing their growth under $\Delta \phi$. This equation looks like a Ward identity obeyed by a pair of marginally deformed CFTs. It would be interesting to independently derive this equation using the pair of CFTs at the boundaries of the wormhole.

\subsection{Wormhole amplitudes between flat boundary metrics} \label{wormflat3d}

We modify the previous computation in order to relate the on-shell gravitational action directly to CFT obervables calculated with respect to a flat background metric. To this end, we first express the hyperbolic metric on the compact surface in the conformal gauge,
\begin{equation}
  \gamma_{ij}dx^idx^j =e^{\Phi(z,\overline{z})}|dz|^2
\end{equation}
where $\Phi$ is the Liouville field which solves the Liouville equation in the absence of sources,
\begin{equation}
  \partial \overline{\partial}\Phi =\frac{e^{\Phi}}{2}
\end{equation}
The metric on the wormhole in hyperbolic slicing coordinates can now be expressed 
\begin{equation}
 ds^2=d\rho^2 + \frac{1}{2}(1+a^2 \cosh(2\rho))e^{\Phi}|dz|^2
\end{equation}
and the Liouville equation for $\Phi$ is consistent with Einstein's equations. The calculation of the on-shell action when $a=1$ i.e for the Maldacena-Maoz wormholes \cite{Maldacena:2004rf} was done in \cite{Chandra:2022bqq}. We generalize that calculation to the deformed wormholes in this section.
To evaluate the regulated on-shell action, we choose a ``wiggly'' cutoff in the $\rho$ coordinate which is dependent on the transverse coordinates at $\rho=\pm \rho_c(z,\overline{z})$ where 
\begin{equation}
    \rho_c = \log (\frac{2}{a \epsilon})-\frac{\Phi(z,\overline{z})}{2}
\end{equation}
so that the metric induced on either boundary to leading order in the cutoff $\epsilon$ is flat, $\frac{1}{\epsilon^2}|dz|^2$ and imporantly independent of $a$. The parameter $a$ enters as part of the boundary condition where we fix the asymptotic values of the scalar field as written above for the Janus solution. Having described the solution, we now want to compute the action. The bulk term evaluates to give
\begin{equation}
    -\frac{1}{16\pi G_N}\int \sqrt{g}(R+2-(\nabla \phi)^2)=\frac{1}{4\pi G_N}\int d^2z \left(\frac{1}{\epsilon^2}+e^\Phi \log(\frac{2}{a\epsilon})-\frac{\Phi}{2}e^\Phi \right )
\end{equation}
The induced metric on the cutoff surfaces is
\begin{equation}
    ds^2_{\text{bdry}}=\left(\frac{1}{\epsilon^2}+\frac{e^\Phi}{2}\right )|dz|^2+\frac{1}{4}(\partial \Phi dz+\overline{\partial}\Phi d\overline{z})^2
\end{equation}
The trace of the extrinsic curvature evaluates to give
\begin{equation}
    K=2+2\epsilon^2 (\partial \overline{\partial}\Phi-\frac{e^\Phi}{2})+O(\epsilon^3)
\end{equation}
Since the Liouville equation is satisfied on-shell, we have $K=2+O(\epsilon^3)$. So, the boundary term evaluates to give
\begin{equation}
    -\frac{1}{8\pi G_N}\int \sqrt{h}(K-1)=-\frac{1}{4\pi G_N}\int d^2 z \left (\frac{1}{\epsilon^2}+\frac{e^\Phi}{2}+\frac{1}{2}\partial \Phi \overline{\partial}\Phi\right )
\end{equation}
The total action is given by
\begin{equation}
    I=\frac{1}{2\pi G_N}\int d^2 z \left ((\partial \Phi \overline{\partial}\Phi +e^\Phi)-\frac{1}{2}\overline{\partial}(\Phi \partial \Phi)-\frac{e^\Phi}{2}(1 + \log (\frac{a \epsilon}{2}))\right )
\end{equation}
For a compact smooth Riemann surface which is boundary-less, the second term vanishes. So, the regulated on-shell action can be expressed as
\begin{equation}
    I(a)=\frac{c}{6}(2S_L)+\frac{c}{3}\chi(1+\log (\frac{ \epsilon}{2}))+\frac{c}{3}\chi \log(a)
\end{equation}
where $S_L$ is the classical Liouville action with the cosmological constant rescaled to $\mu=\frac{1}{4\pi b^2}$ given by
\begin{equation}
  S_L=\frac{1}{4\pi}\int d^2 z \left (\partial \Phi \overline{\partial} \Phi+e^{\Phi} \right )
\end{equation}
Now, we make a couple of comments about this calculation.
Firstly, notice that the Liouville part of the action which explicitly depends on the moduli of the compact surface factorises from the part of the action sensitive to the difference in marginal couplings which is affected only by the topology of the compact surface and has no explicit moduli dependence. This means that the difference in the regulated actions of the deformed and undeformed wormholes remains unchanged from the previous case where they were calculated with a hyperbolic metric on the two boundaries,
\begin{equation}
  ( I(a)- I(1))\bigg |_{|dz|^2} =( I(a)- I(1))\bigg |_{e^{\Phi}|dz|^2 }=\frac{c}{3}\chi \log(a)
\end{equation}
Secondly, note that the Brown-York stress tensor on either boundary is invariant under the deformation so it is traceless and factorises holomorphically into the Liouville stress tensors for the left and right movers,
\begin{equation}
    T(z)=\frac{c}{6}T^\Phi(z), \qquad T^{\Phi}(z)=\frac{1}{2}\partial^2 \Phi -\frac{1}{4}(\partial \Phi)^2
\end{equation}
with a corresponding expression for the antiholomorphic component. It is important to emphasize that the stress tensor continuing to be traceless shows that the Janus-like deformation is marginal for both the CFTs on a flat background metric.

\subsection{Deformed OPE statistics of local operators in 2d CFT} \label{OPE}

We interpret the wormhole amplitudes computed so far in terms of averages of observables between the two copies of CFT on either boundary deformed by different marginal couplings to the scalar operator dual to the dilaton. If we denote the partition function of CFT deformed by a marginal coupling $\phi$ on the compact hyperbolic surface $\Sigma$ calculated with a flat background metric by $Z[\Sigma]^\phi$, then we have the following result expressing the averages of these partition functions in terms of the partition functions of two-boundary wormholes with the topology of $\Sigma \times \mathbb{R}$ at large-$c$:
\begin{equation} \label{corr3d}
  \overline{Z[\Sigma]^{\phi_0+\frac{\Delta \phi}{\sqrt{2}}}Z[\Sigma]^{\phi_0-\frac{\Delta \phi}{\sqrt{2}}}}=(\cosh(\Delta \phi))^{\frac{c\chi(\Sigma)}{6}}\overline{Z[\Sigma]Z[\Sigma]}
\end{equation}
where $\chi(\Sigma)$ is the Euler characteristic of the surface. Since $\chi(\Sigma)<0$, this result shows that the correlation between the pair of CFT partition functions decreases monotonically with the difference in the marginal couplings. Note that this formula was derived assuming $\Sigma$ is smooth. However, it can be readily generalised to include surfaces punctured by conical defects with the replacement of $\chi(\Sigma)$ by its generalisation to conifolds. For example, a sphere punctured by $n$ conical defects with opening angles $2\pi(1-2\eta_i)$ with $\eta_i \in (0,\frac{1}{2})$ has 
\begin{equation}
   \chi(\Sigma_{0,n})=2(1-\sum_{i=1}^n \eta_i)
\end{equation}
The condition to put a hyperbolic metric is that the above expression for $\chi$ is negative. We show that this expectation is consistent with the calculation of the gravitational action on $\Sigma_{0,n} \times \mathbb{R}$ with different asymptotic values for the dilaton on the two boundaries. To this end, we modify the gravitational action in \eqref{action3d} to include the worldlines of massive point particles,
\begin{equation}
   I=-\frac{1}{16\pi G_N}\int \sqrt{g}(R+2-(\nabla \phi)^2)-\frac{1}{8\pi G_N}\int\sqrt{h}(K-1)-\sum_{i=1}^n m_i \int ds_i
\end{equation}
with the masses tuned to the conical deficit angles by $m_i=\frac{c\eta_i}{3}$. It is important to note that the dilaton is not coupled to the matter worldlines. Computing the on-shell action for the wormhole in the same way as done in section \ref{wormflat3d} and by suitably regulating the defects as described in \cite{Chandra:2022bqq}, we get
\begin{equation}
  I(a)=\frac{c}{3}\left (S_L-2\log R +2\sum_{i=1}^n \eta_i^2 \log \epsilon_i-2(1+\log (\frac{\epsilon}{2}))(\sum_{i=1}^n \eta_i-1)\right )-\frac{2c}{3}(\sum_{i=1}^n \eta_i-1)\log(a)
\end{equation}
Here, $S_L$ is the semiclassically renormalised Liouville action on $\Sigma_{0,n}$ written down for instance in \cite{Harlow:2011ny}; $\epsilon_i$ are the sizes of the small disks excised around the defects and $R$ is the large radius cutoff on the punctured sphere. The important piece is the second term which shows that the difference in the regulated actions of the deformed and undeformed wormholes is given by
\begin{equation}
  I(a)-I(1)=\frac{c}{6}\chi(\Sigma_{0,n})\log (a)
\end{equation}
therefore proving our expectation. 

A particularly nice consequence of the above result when applied to $\Sigma =\Sigma_{0,3}$ is the following universal large-$c$ expression for the variance of the OPE coefficients of local operators between CFTs deformed by different marginal couplings,
\begin{equation} \label{OPEdef}
    \overline{c_{ijk}^{\phi_0+\frac{\Delta \phi}{\sqrt{2}}}c_{ijk}^{\phi_0-\frac{\Delta \phi}{\sqrt{2}}}}=|C_0(h_i,h_j,h_k)|^2 \left(\cosh(\Delta \phi) \right )^{-\frac{c}{3}(\eta_i+\eta_j+\eta_k-1)} 
\end{equation}
with the OPE coefficients between the heavy sub-threshold scalar operators with conformal weights parametrized as $h=\frac{c}{6}\eta(1-\eta)$ with $\eta \in (0,\frac{1}{2})$ defined by
\begin{equation} \label{VirETH}
 c_{ijk}^{\phi}\equiv \langle O_i(0) O_j(1) O_k(\infty) \exp(-\phi \int d^2z \mathcal{O}(z,\overline{z})) \rangle
\end{equation}
Here, $\mathcal{O}$ is the marginal scalar dual to the dilaton and is integrated over $\mathbb{C}\backslash \{0,1,\infty\}$ and the heavy operators are normalized such that in the undeformed CFT, the variance of their OPE coefficients is set by a smooth function of the conformal weights usually called $C_0$ determined using the crossing tranformation of the identity block \cite{Collier:2019weq},
\begin{equation}
    \overline{|c_{ijk}|^2}=|C_0(h_i,h_j,h_k)|^2
\end{equation}
We have also used conformal invariance of the marginally deformed theory to set the locations of the heavy operators to the standard reference locations at $0,1,\infty$. Extrapolating between \eqref{OPEdef} and \eqref{corr3d} which was derived for smooth $\Sigma$ and demanding holomorphic factorization, we propose the following large-$c$ expression for the variance of the OPE coefficients between any three heavy local operators whose conformal weights scale linearly with $c$,
\begin{align} \label{OPElocal}
    \overline{c_{ijk}^{\phi_0+\frac{\Delta \phi}{\sqrt{2}}}c_{ijk}^{\phi_0-\frac{\Delta \phi}{\sqrt{2}}}}=|C_0(h_i,h_j,h_k)|^2 \left(\cosh(\Delta \phi) \right )^{-\frac{c}{6}(\text{Re}(\eta_i+\eta_j+\eta_k)-1)} \left(\cosh(\Delta \phi) \right )^{-\frac{c}{6}(\text{Re}(\overline{\eta}_i+\overline{\eta}_j+\overline{\eta}_k)-1)}
\end{align}
where the conformal weights of the operators are parametrised as
\begin{equation}
    h=\frac{c}{6}\eta(1-\eta), \qquad \overline{h}=\frac{c}{6}\overline{\eta}(1-\overline{\eta})
\end{equation}
For operators above the black hole threshold, $\eta = \frac{1}{2}+i\mathbb{R}$ and for operators below the black hole threshold, $\eta \in (0,\frac{1}{2})$. It is worth emphasizing that if we restrict to operators above the black hole threshold i.e with conformal weights $h,\overline{h}>\frac{c}{24}$, then the deformed variance proposed in \eqref{OPElocal} takes the following universal form,
\begin{equation} \label{varlocheavy}
   \overline{c_{ijk}^{\phi_0+\frac{\Delta \phi}{\sqrt{2}}}c_{ijk}^{\phi_0-\frac{\Delta \phi}{\sqrt{2}}}}=|C_0(h_i,h_j,h_k)|^2 \left(\cosh(\Delta \phi) \right )^{-\frac{c}{6}} 
\end{equation}
suggesting that the deformation $\Delta \phi$ corresponding to the difference in marginal couplings does not distinguish between operators above the black hole threshold at large $c$. The proposal \eqref{varlocheavy} is consistent with the prediction \eqref{corr3d} from smooth wormhole amplitudes which can be easily verified by expanding the partition functions $Z[\Sigma]$ in \eqref{corr3d} using a suitable channel and performing the Wick contractions of the OPE coefficients using \eqref{varlocheavy}. A more direct check of \eqref{varlocheavy} is the consistency with the wormhole amplitudes that directly compute the variance between the OPE coefficients of operators above the black hole threshold constructed in \cite{Abajian:2023bqv}.

 In conclusion, the Gaussian statistics with variance given by \eqref{OPElocal} is consistent with the partition fuctions on all two-boundary wormholes with the topology of $\Sigma_{g,n}\times \mathbb{R}$ at large $c$. Consistency of these statistics with the eigenstate thermalization hypothesis (ETH) \cite{Srednicki_1994} would serve as a precision test for averaging in AdS/CFT analogous to the one suggested in \cite{Cotler:2022rud} for the spectral form factor where the consistency being tested is between the partition functions of wormholes with the topology of $T^2 \times \mathbb{R}$ (and their higher dimensional analogues \cite{Cotler_2021,Cotler:2022rud}) and different couplings on the two boundaries, with the prediction from random matrix theory\footnote{I thank Tom Hartman for suggesting the interpretation of the result in \eqref{OPElocal} in terms of a precision test for ETH in analogy with the precision test of \cite{Cotler:2022rud}.}. 

\subsection{Deformed statistics of matrix elements of line defects in 2d CFT} \label{Linedef}

We can generalize the discussion of OPE statistics of local operators to compute the statistics of the matrix elements of line defects. Specifically, we work with line defects $D_\Sigma$ dual to spherically symmetric thin shells of dust particles in 3d gravity constructed in \cite{Chandra:2024vhm} builiding on the proposal of \cite{Anous:2016kss}. These defects are constructed from continuous products of probe local operators. For the case where the defect is placed along a circle, 
\begin{equation} \label{linedef}
  D_\Sigma= \prod_{j=1}^N \mathcal{O}_j\left (\frac{2\pi j}{N} \right)
\end{equation}
the probe local operators $\mathcal{O}_j$ are uniformly distributed around the circle with the number of operators $N \sim \sqrt{c}$ and their conformal dimensions $h \sim \sqrt{c}$. The mass $m$ of the defect is defined as $m \equiv \frac{12Nh}{c}$. In the dual gravitational theory, this corresponds to the mass of the shell.
To compute the variance of the matrix elements of this line defect holographically, we work with the gravitational theory described by
\begin{equation}
  I=-\frac{1}{16\pi G_N}\int \sqrt{g}(R+2-(\nabla \phi)^2)-\frac{1}{8\pi G_N}\int\sqrt{h}(K-1)-\sum m_i \int ds_i
\end{equation}
where $m_i$ are the masses of the corresponding light probe particles. Specifically, we look for two-boundary wormhole solutions with a single spherically symmetric thin shell of mass $m$ and wordlines of two heavy point particles with deficit angles parametrized by $\eta_i$ and $\eta_j$ going through it. We express the metric on this wormhole as
\begin{equation}
 ds^2=d\rho^2 +\frac{1}{2}(1+a^2\cosh(2\rho))e^{\Phi(y)}(dy^2+d\theta^2) \qquad \qquad y\in (-\infty, \infty), \theta \in (0,2\pi)
\end{equation}
The Liouville solution $\Phi(y)$ corresponds to a line defect placed along the circle $y=0$ and two conical defects at $y \to \pm \infty$ in the cylinder frame was constructed in \cite{Chandra:2024vhm} and reviewed in the appendix \ref{thinshellapp}. The wormhole solves the Israel junction conditions for gluing 3d geometries across a thin shell of dust \cite{Israel:1966rt},
\begin{equation}
  \Delta K_{ij}-\Delta K h_{ij}= \sigma(\rho) U_i U_j
\end{equation}
where $U=\partial_\rho$ is the normalised velocity field for the particles constituting the shell and the energy density of the shell is
\begin{equation}
 \sigma(\rho)=\frac{m}{e^{\frac{\Phi_0}{2}}\sqrt{\frac{1}{2}(1+a^2\cosh(2\rho))}}
\end{equation}
 where $\Phi_0$ is the value of the Liouville field at the transverse location of the shell i.e, at $y=0$ in these coordinates.
 By computing the on-shell action for this wormhole generalising the calculation done in \cite{Chandra:2024vhm} for the undeformed i.e, $a=1$ wormhole, we find that the difference in actions between the deformed and undeformed wormholes is finite just like in all the earlier examples and is given by
\begin{equation}
  I(a)-I(1)=-\frac{2c}{3}(\eta_i+\eta_j+\frac{m}{2}-1)\log(a)
\end{equation}
Thus, the variance of the matrix elements of the line defect between heavy sub-threshold scalar operators, between the CFTs with different marginal couplings is given by
\begin{equation} \label{defvarlight}
   \overline{\bra{i}D_\Sigma \ket{j}^{\phi_0+\frac{\Delta \phi}{\sqrt{2}}}\bra{j}D_\Sigma^\dagger \ket{i}^{\phi_0-\frac{\Delta \phi}{\sqrt{2}}}}=|C_0^D(h_i,h_j;m)|^2 \left(\cosh(\Delta \phi) \right )^{-\frac{c}{3}(\eta_i+\eta_j+\frac{m}{2}-1)}
\end{equation}
where $h=\frac{c}{6}\eta(1-\eta)$ are the conformal dimensions of the heavy scalar operators. The matrix elements of the line defects in the deformed theory defined as
\begin{equation}
   \bra{i}D_\Sigma \ket{j}^\phi \equiv \langle O_i(0)D_\Sigma O_j(\infty) \exp(-\phi \int d^2z \mathcal{O}(z,\overline{z}))\rangle
\end{equation}
with the operator normalisations chosen so that in the undeformed theory, the variance of the matrix elements is set by the smooth function $C_0^D$ of the defect mass and operator weights whose explicit form was derived by studying crossing of the infinite-point identity block at large $c$ in \cite{Chandra:2024vhm},
\begin{equation}
     \overline{|\bra{i}D_\Sigma \ket{j}|^2}=|C_0^D(h_i,h_j;m)|^2
\end{equation}
More generally, we propose the following correction to the variance of the matrix elements of the line defects between the CFTs with different marginal couplings,
\begin{multline}
    \overline{\bra{i}D_\Sigma \ket{j}^{\phi_0+\frac{\Delta \phi}{\sqrt{2}}}\bra{j}D_\Sigma^\dagger \ket{i}^{\phi_0-\frac{\Delta \phi}{\sqrt{2}}}}=|C_0^D(h_i,h_j;m)|^2 \left(\cosh(\Delta \phi) \right )^{-\frac{c}{6}(\text{Re}(\eta_i+\eta_j)+\frac{m}{2}-1)}\\ \left(\cosh(\Delta \phi) \right )^{-\frac{c}{6}(\text{Re}(\overline{\eta}_i+\overline{\eta}_j)+\frac{m}{2}-1)}
\end{multline}
where the conformal weights of the local operators are parametrised as
\begin{equation}
    h=\frac{c}{6}\eta(1-\eta), \qquad \overline{h}=\frac{c}{6}\overline{\eta}(1-\overline{\eta})
\end{equation}
In particular, the above proposal implies that when the external operators are above the black hole threshold, the variance takes the form,
\begin{equation} \label{defvarshell}
   \overline{\bra{i}D_\Sigma \ket{j}^{\phi_0+\frac{\Delta \phi}{\sqrt{2}}}\bra{j}D_\Sigma^\dagger \ket{i}^{\phi_0-\frac{\Delta \phi}{\sqrt{2}}}}=|C_0^D(h_i,h_j;m)|^2 (\cosh(\Delta \phi))^{-\frac{c}{6}m}
\end{equation}
This result is consistent with the wormhole amplitudes where the shell(s) propagates between higher genus Riemann surfaces on the boundaries. For example, consider the family of wormholes which compute the variance of the thermal two-point function of the line defect. The Riemann surfaces on the boundaries of these wormholes are tori backreacted on by a pair of parallel line defects. By turning on the Janus deformation, the wormhole amplitude relates the deformed variance to the undeformed variance by
\begin{equation} \label{torustwoshell}
  \overline{\langle D^\dagger(\tau_0) D(-\tau_0) \rangle_\beta^{\phi_0+\frac{\Delta \phi}{\sqrt{2}}}\langle D^\dagger(\tau_0) D(-\tau_0) \rangle_\beta^{\phi_0-\frac{\Delta \phi}{\sqrt{2}}}}=\overline{\left |\langle D^\dagger(\tau_0) D(-\tau_0) \rangle_\beta \right |^2} (\cosh(\Delta \phi))^{-\frac{c}{3}m}
\end{equation}
where the thermal two-point function is defined by
\begin{equation} 
  \langle D^\dagger(\tau_0) D(-\tau_0) \rangle_\beta \equiv \text{Tr}\left (e^{-(\beta-2\tau_0) H} D_\Sigma^\dagger e^{-2\tau_0 H} D_\Sigma \right )
\end{equation}
and a corresponding expression in the marginally deformed theory. We emphasize that \eqref{torustwoshell} is a prediction from the gravity calculation. To arrive at this result, we made use of the fact that the generalized Euler characteristic calculated from the area of the torus deformed by the line defects is given by
\begin{equation} \label{torchi}
  \chi \equiv -\frac{1}{2\pi}\int_{T^2} d^2 z e^{\Phi(z,\overline{z})}=-2m
\end{equation}
where $\Phi$ is the Liouville field on this surface written down in \cite{Chandra:2024vhm} reviewed in Appendix \ref{Liouvtor} where we also derive \eqref{torchi}. Note that the result \eqref{torchi} is independent of the location $\tau_0$ of the line defect. There are two Wick contractions between the four matrix elements of $D_\Sigma$ that show up in the expansion of the product of thermal two-point functions on the LHS of \eqref{torustwoshell}. Performing these contractions using the proposal \eqref{defvarshell}, the result is consistent with the prediction from the wormhole amplitude \eqref{torustwoshell}.

A special case of the above result in \eqref{defvarlight} is the variance of the vacuum expectation value sometimes referred to as the $g$-function \cite{Affleck:1991tk,Cuomo:2021rkm} of the line defect,
\begin{equation}
     \overline{\langle D_\Sigma \rangle^{\phi_0+\frac{\Delta \phi}{\sqrt{2}}}\langle D_\Sigma^\dagger \rangle^{\phi_0-\frac{\Delta \phi}{\sqrt{2}}}}=\overline{|\langle D_\Sigma \rangle|^2}(\cosh(\Delta \phi))^{-\frac{c}{6}(m-2)}
\end{equation}
where the variance of the $g$-function in the undeformed theory was calculated in \cite{Chandra:2024vhm} and reviewed in Appendix \ref{thinshellapp},
\begin{equation}
    \overline{|\langle D_\Sigma \rangle|^2}= \left(e^{-m}(m-2)^{m/2-1}(m+2)^{m/2+1}\right)^{\frac{c}{3}}
\end{equation}
If the line defect $D_\Sigma$ is charged under a global symmetry of the CFT which is not gauged in the bulk then the above variance gives an estimation of the amount of global charge violation described by a wormhole amplitude as observed in \cite{Chandra:2024vhm}\footnote{See also \cite{Hsin_2021,Chen_2021,Bah_2023} for recent works relating global charge violating amplitudes and wormholes.}. Now, the current expression for the variance between marginally deformed theories quantifies how this global charge violating amplitude decays as the volume of the wormhole grows with the difference in marginal couplings.

\subsection{Dilaton correlators from wormhole amplitudes} \label{dil3d}

The two-boundary wormholes discussed so far are gravitational contributions to the averaged product of CFT observables on either boundary which is a compact hyperbolic surface $\Sigma_{g,n}$.
\begin{equation} \label{dilgen3d}
    Z_{\text{WH}}[\phi_+,\phi_-;\Sigma_{g,n}] = \overline{\left \langle \exp \left(-\phi_+\int_{\Sigma_{g,n}} d^2 x \sqrt{g}  \mathcal{O}(x)\right) \right \rangle \left \langle \exp \left(-\phi_-\int_{\Sigma_{g,n}} d^2 x \sqrt{g} \mathcal{O}(x)\right) \right \rangle}
\end{equation}
The sources $\phi_\pm$ in the CFT path integral for the marginal scalar $\mathcal{O}$ (has conformal weights $h=\overline{h}=1$) dual to the dilaton are the asymptotic values of dilaton on either boundary. On the RHS, the product of CFT observables is calculated placing the CFTs on the compact Riemann surfaces. The domain of the integral in \eqref{dilgen3d} could be chosen to be a fundamental domain for the quotient of the upper half plane by a discrete subgroup of PSL$(2,\mathbb{R})$ called a Fuchsian group, corresponding to the compact surface $\Sigma_{g,n}$. The product of CFT observables is averaged over a small window around the marginal couplings. It is important to note that this product involves obervables in different CFTs. The fact that this product receives a wormhole contribution means that there is ensemble averaging even for observables involving very light operators (whose conformal dimension doesn't scale with $c$ as in the current example) when the CFTs are placed on higher genus surfaces.

Applying the GKPW dictionary \cite{Gubser:1998bc,Witten:1998qj} which relates operator sources in the CFT to asymptotic values of the corresponding bulk fields, we treat the wormhole amplitude as a generating functional for averaged dilaton correlators. In that case, $\log (Z_{\text{WH}})$ is a generating functional for averaged connected correlators. Since we have explicitly evaluated the wormhole amplitudes only for dilaton profiles which are uniform over $\Sigma_{g,n}$ i.e, the dilaton sources are uniform along the boundaries, we can only extract integrated correlators by differentiating the wormhole amplitudes with respect to these sources.  The averaged integrated scalar one-point functions on either boundary are given by,
\begin{equation}
 \frac{\overline{ \left \langle \int_{\Sigma_{g,n}} d^2 x \sqrt{g}  \mathcal{O}(x)\exp \left(-\phi_\pm\int_{\Sigma_{g,n}}  d^2 x \sqrt{g}  \mathcal{O}(x)\right) \right \rangle   \left \langle  \exp \left(-\phi_\mp\int_{\Sigma_{g,n}}  d^2 x \sqrt{g}  \mathcal{O}(x)\right) \right \rangle }}{\overline{\left \langle \exp \left(-\phi_+\int_{\Sigma_{g,n}}  d^2 x \sqrt{g}  \mathcal{O}(x)\right) \right \rangle \left \langle \exp \left(-\phi_-\int_{\Sigma_{g,n}}  d^2 x \sqrt{g}  \mathcal{O}(x)\right) \right \rangle}}= -\frac{\partial \log (Z_{\text{WH}})}{\partial \phi_\pm}
\end{equation}
We showed that the wormhole amplitude rescaled by the partition function of the undeformed wormhole at large-$c$ is given by
\begin{equation}
    Z_{\text{WH}}[\phi_+,\phi_-;\Sigma_{g,n}]=(\cosh(\Delta \phi))^{\frac{c \chi(\Sigma_{g,n})}{6}},\qquad \Delta \phi=\frac{\phi_+-\phi_-}{\sqrt{2}}
\end{equation}
The averaged integrated one-point functions denoted as $G^{\pm}_1$ on either boundary are therefore given by
\begin{equation}
   G_1^{\pm}(\Delta \phi)=\pm \frac{c|\chi|}{6\sqrt{2}}\tanh(\Delta \phi)
\end{equation}
It is not expected that the one-point function is homogeneous since we are not computing it in the ground state of the CFT.
The fact that the integrated one-point functions are equal in magnitude but opposite in sign is expected since the wormhole amplitude only depends on the difference in the asymptotic values of the dilaton. We see that the magnitude of the integrated one-point function increases monotonically with the difference in couplings and is bounded from above by $\frac{c|\chi|}{6\sqrt{2}}$. 

Now, we turn to computing the integrated two-point functions. They are of two types: When the insertions are on the same boundary, we call them self-correlators denoted $G^\pm_2$ and when they are on different boundaries, we call them cross-correlators denoted $G_{1,1}$. The connected self-correlators are defined as 
\begin{align}
  & G^\pm_2=-(G_1^\pm)^2+\notag \\ & \frac{\overline{ \left \langle \int_{\Sigma_{g,n}} d^2 x d^2 y\sqrt{g(x)g(y)}  \mathcal{O}(x)\mathcal{O}(y)\exp \left(-\phi_\pm\int_{\Sigma_{g,n}}  d^2 x \sqrt{g}  \mathcal{O}(x)\right) \right \rangle   \left \langle  \exp \left(-\phi_\mp\int_{\Sigma_{g,n}}  d^2 x \sqrt{g}  \mathcal{O}(x)\right) \right \rangle }}{\overline{\left \langle \exp \left(-\phi_+\int_{\Sigma_{g,n}}  d^2 x \sqrt{g}  \mathcal{O}(x)\right) \right \rangle \left \langle \exp \left(-\phi_-\int_{\Sigma_{g,n}}  d^2 x \sqrt{g}  \mathcal{O}(x)\right) \right \rangle}}
\end{align}
while the connected cross-correlators are defined as 
\begin{align}
   & G_{1,1}=-G_1^+G_1^-+\notag \\ & \frac{\overline{ \left \langle \int_{\Sigma_{g,n}} d^2 x \sqrt{g}  \mathcal{O}(x)\exp \left(-\phi_+\int_{\Sigma_{g,n}}  d^2 x \sqrt{g}  \mathcal{O}(x)\right) \right \rangle   \left \langle \int_{\Sigma_{g,n}} d^2 x \sqrt{g}  \mathcal{O}(x)  \exp \left(-\phi_-\int_{\Sigma_{g,n}}  d^2 x \sqrt{g}  \mathcal{O}(x)\right) \right \rangle }}{\overline{\left \langle \exp \left(-\phi_+\int_{\Sigma_{g,n}}  d^2 x \sqrt{g}  \mathcal{O}(x)\right) \right \rangle \left \langle \exp \left(-\phi_-\int_{\Sigma_{g,n}}  d^2 x \sqrt{g}  \mathcal{O}(x)\right) \right \rangle}}
\end{align}
For the present case, they evaluate to give
\begin{equation} \label{twopoint3d}
  \begin{split}
     & G_2^\pm (\Delta \phi) = \frac{\partial^2 \log Z_{\text{WH}}}{\partial \phi_\pm^2}=-\frac{ c|\chi|}{12}\text{sech}^2(\Delta \phi)\\
     & G_{1,1} (\Delta \phi)=  \frac{\partial^2 \log Z_{\text{WH}}}{\partial \phi_+ \partial  \phi_-}=\frac{c|\chi|}{12}\text{sech}^2(\Delta \phi)
  \end{split}
\end{equation}
Not surprisingly, since the wormhole amplitude only depends on the difference in marginal couplings, we see that the self-correlators and cross-correlators are equal in magnitude but oppposite in sign. The self-correlators are negative while the cross-correlators are positive. We notice that when the deformation is turned off, these correlators still take non-zero values,
\begin{equation} \label{max3d}
   G_{1,1}=- G_2^\pm =\frac{c |\chi|}{12} \qquad \Delta \phi=0
\end{equation}
These are values of the integrated dilaton correlators calculated in the background of the undeformed Maldacena-Maoz wormholes.
The magnitudes of the self- and cross-correlators decay monotonically with the deformation below the maximum value \eqref{max3d}. This decay of the cross-correlator corroborates the observation made earlier that the CFTs are getting increasingly decorrelated as the difference in marginal couplings increases.

\section{Wormholes in higher dimensional holographic RG flows} \label{higherdwormholes}

In this section, we generalize the analysis done in the previous section to higher dimensions using the higher dimensional analogues of the Janus domain wall solutions in Einstein-dilaton gravity. At a qualitative level, the results of this section agree with those in the previous section although the analysis is more involved. We start by reviewing the higher dimensional Janus solutions. We then construct two-boundary wormholes using the Janus solutions, compute their volumes and prove that they increase monotonically with the difference in the asymptotic dilaton values, and show that the wormhole amplitudes between flat or hyperbolic boundary metrics can be expressed in terms of the wormhole volumes. Finally, we use the wormhole amplitudes to compute the integrated dilaton correlators.

\subsection{Review: Janus domain walls in higher dimensions}

The Janus-like domain wall solutions of Einstein-dilaton gravity also exist in higher dimensions. In $5d$, they were constructed by a dimensional reduction of $10d$ domain wall solutions of IIB supergravity on $AdS_5\times S^5$ in \cite{Bak:2003jk} and the stability of these solutions was analyzed in \cite{Freedman:2003ax}. The field theory interpretation for these solutions in terms of a conformal interface in $\mathcal{N}=4$ SYM across which $g^2_{YM}$ jumps was provided in \cite{Clark:2004sb}. The correlation functions in these domain wall solutions were calculated holographically in general dimensions in \cite{Papadimitriou:2004rz}. We review these solutions in this section very closely following the analysis in \cite{Papadimitriou:2004rz} and adding a few details where necessary. The action for Einstein-dilaton gravity in Euclidean signature in $(d+1)$-bulk dimensions is
\begin{equation}
    I=-\frac{1}{16\pi G_N}\int \sqrt{g}(R+d(d-1)-(\nabla \phi)^2)-\frac{1}{8\pi G_N}\int \sqrt{h}K+I_{ct}[h]
\end{equation}
where $I_{ct}$ denotes counter-terms needed to make the action finite and is a functional only of the boundary metric. We have set the AdS scale to $1$ and the cosmological constant to $\Lambda=-\frac{d(d-1)}{2}$.
Varying the action, Einstein's equations take the form,
\begin{equation}
    R_{\mu \nu}-\frac{1}{2}g_{\mu \nu}R-\frac{d(d-1)}{2}g_{\mu \nu}=\partial_\mu \phi \partial_\nu \phi -\frac{1}{2}g_{\mu \nu}(\nabla \phi)^2
\end{equation}
Taking the trace of the above equation gives,
\begin{equation}
    R=-d(d+1)+(\nabla \phi)^2
\end{equation}
using which the Einstein equation can be rewritten as
\begin{equation}
    R_{\mu \nu}+d g_{\mu\nu}=\partial_\mu \phi \partial_\nu \phi
\end{equation}

The metric on these domain wall solutions takes a simple closed form when expressed in terms of the new radial coodinate $u$\footnote{We can relate the form of the metric in \eqref{Janushigherd} to the more familiar form of the domain wall metric 
\begin{equation} \label{domainwall}
   ds^2=dr^2+e^{2A(r)}dH_d^2
\end{equation}
with $r \in (0,\infty)$ using the coordinate tranformation relating $u$ to the scale factor by $u=e^{-2A(r)}$. Einstein's equations express the radial evolution of the scale factor in terms of $u$ as
\begin{equation}
  \Dot{A}^2=1-u+bu^d
\end{equation}
The form of the metric in \eqref{domainwall} can be analytically extended to cover the full domain wall solution by continuing the function $A(r)$ to $r<0$ as an even function, $A(r)=A(-r)$. This obervation also generalizes to the two-boundary wormhole solutions that we construct subsequently i.e the coordinate tranformation from \eqref{Janushigherd} to \eqref{domainwall} allows us to cover the full wormhole solution. },
\begin{equation} \label{Janushigherd}
    ds^2=\frac{du^2}{4u^2(1-u+bu^d)}+\frac{1}{u}dH_d^2
\end{equation}
where $dH_d^2$ is the Poincare metric on upper half space $H_d$. This coordinate system covers half of the solution and the range of the $u$ coordinate is $(0,u_0)$ where $u_0$ is the smallest positive root of the equation, $1-u+b u^d=0$. Clearly, we see that since the deformation parameter $b$ is taken to be non-negative, $u_0\geq 1$ with the inequality getting saturated when $b=0$. The solution is non-singular provided
\begin{equation}
    0 \leq b <b_0 = \frac{1}{d}\left(\frac{d-1}{d}\right)^{d-1}
\end{equation}
The upper bound comes from requiring that the polynomial $1-u+b u^d$ has a double root at $u_0$ with
\begin{equation}
    u_0(b_0)=\frac{d}{d-1}
\end{equation}
For a general value of $b$, we can write a closed form expression for $u_0$ in terms of generalized hypergeometric functions\footnote{This series for $u_0$ can be derived by repeatedly differentiating $1-u_0+bu_0^d=0$ with respect to $b$ to obtain all the derivatives at $b=0$, $u_0^{(n)}(0)$.},
\begin{multline}
     u_0(b)=\sum_{n=0}^{\infty} \frac{\Gamma(nd+1)}{\Gamma(n(d-1)+2)}\frac{b^n}{n!} \\= {}_{d-1} F_{d-2}\left [\left (\frac{1}{d},\frac{2}{d},\dots,\frac{d-1}{d} \right ),\left (\frac{2}{d-1},\frac{3}{d-1},\dots,\frac{d-2}{d-1},\frac{d}{d-1}\right ),\frac{b}{b_0}\right ]
\end{multline}
The generalized hypergeometric function above takes a simple form in $d=2$ and $d=3$,
\begin{equation}
    \begin{split}
        & u_0(b)=\frac{1}{2b}(1-\sqrt{1-4b}) \qquad d=2\\
        & u_0(b)=\frac{2}{\sqrt{3b}}\sin(\frac{1}{3}\sin^{-1}(\frac{3}{2}\sqrt{3b})) \qquad d=3
    \end{split}
\end{equation}
Since $u_0$ has a convergent Taylor series expansion for $b<b_0$ with positive coefficients, it increases montonically with $b$.
When $b=0$, we can transform the above coordinates to the more familiar hyperbolic slicing coordinates by writing $\frac{1}{u}=\cosh^2(\rho)$ to get
\begin{equation}
    ds^2=d\rho^2+\cosh^2(\rho)\gamma_{ij}dx^idx^j \qquad b=0
\end{equation}
and analytically continue the above metric to $\rho \in \mathbb{R}$ to cover the full solution. When $d=2$ i.e, for the Janus solution in AdS$_3$, we can relate this coordinate system to the one discussed in the previous section by
\begin{equation}
    \frac{1}{u}=\frac{1+a^2 \cosh(2\rho)}{2} \qquad b=\frac{1-a^4}{4}
\end{equation}
For this case, $u_0=\frac{2}{1+a^2}$. Since $a \leq 1$, we see that $u_0 \geq 1$. 

The equation of motion for the dilaton is the massless Klein-Gordon equation,
\begin{equation}
    \Box \phi=\frac{1}{\sqrt{g}}(\sqrt{g}g^{\mu \nu}\partial_\mu \phi)_{,\nu}=0
\end{equation}
Substituting the ansatz $\phi=\phi(u)$, the equation of motion can be integrated once to give
\begin{equation}
    \partial_u \phi = \frac{c}{2}\frac{u^{\frac{d}{2}-1}}{\sqrt{1-u+bu^d}}
\end{equation}
where the integration constant $c$ is related to the parameter $b$ using the $uu$ component of Einstein's equation $b=\frac{c^2}{d(d-1)}$. We denote the asymptotic values of the dilaton obtained by integrating the above equation by $\phi_0 \pm \frac{\Delta \phi}{2}$. For purposes of this work, the difference in asymptotic values $\Delta \phi$ plays an important role and we have evaluated this difference explicitly in Appendix \ref{dilatonapp} to be given in terms of the generalized hypergeometric functions,
\begin{multline} \label{Deltaphi}
    \Delta \phi =a_d\sqrt{bd(d-1)}\frac{(d-2)!!}{(d-1)!!} {}_d F_{d-1}\bigg [\left ( \frac{d}{2d},\frac{d+2}{2d},\dots, \frac{3d-2}{2d}\right),\\ \left ( \frac{d+1}{2(d-1)},\frac{d+3}{2(d-1)},\dots, \frac{3d-5}{2(d-1)},\frac{3}{2}\right),\frac{b}{b_0(d)}\bigg]
\end{multline}
where the coefficient $a_d$ takes different values for even and odd $d$,
\begin{equation}
    a_d=
    \begin{cases}
         2 \qquad & d \in 2\mathbb{Z}_+ \\
         \pi \qquad & d \in 2\mathbb{Z}_+ +1
    \end{cases}
\end{equation}
For concreteness, we write down \eqref{Deltaphi} for some special values of $d$. In AdS$_4$, it takes the form
\begin{equation}
  \Delta \phi= \pi \sqrt{\frac{3b}{2}} {}_3 F_2 \bigg [\left (\frac{1}{2},\frac{5}{6},\frac{7}{6}\right),\left(1,\frac{3}{2}\right ),\frac{27b}{4}\bigg]
\end{equation}
and in AdS$_5$, it takes the form
\begin{equation} \label{ads5deltaphi}
  \Delta \phi = 8 \sqrt{\frac{b}{3}} {}_4 F_3 \bigg [ \left (\frac{1}{2},\frac{3}{4},1,\frac{5}{4} \right ), \left( \frac{5}{6},\frac{7}{6},\frac{3}{2}\right ),\frac{256b}{27}\bigg ]
\end{equation}
For the two-boundary wormhole solutions described in the next section, $\Delta \phi$ is the difference in the values of the dilaton on the two boundaries of the wormhole. The important features of this result that we shall be using in the subsequent analysis is that the function $G(b)\equiv \frac{\Delta \phi}{\sqrt{bd(d-1)}}$ is non-negative, analytic in $b$ for $0\leq b<b_0$ and monotonically increasing with $b$. As a consequence of this, $\Delta \phi$ is also non-negative and increases montonically with $b$. 

\subsection{Wormhole amplitudes between hyperbolic boundary metrics}

Now, we describe two-boundary wormhole solutions which are locally described by the same metric and dilaton profile as the single-boundary domain wall solutions discussed above. The radial slices and hence the boundaries are quotients of $H_d$ by discrete subgroups of its isometry group which are compact and have finite volume. The metric on these wormhole solutions in hyperbolic slicing can thus be written as
\begin{equation}
    ds^2=\frac{du^2}{4u^2(1-u+bu^d)}+\frac{1}{u}\gamma_{ij}dx^idx^j
\end{equation}
where $\gamma_{ij}dx^idx^j$ is the hyperbolic metric on the $d$-dimensional compact hyperbolic space. As discussed in the previous section, this coordinate system coveres half of the wormhole with the boundary at $u=0$ and the neck at $u=u_0$.
Having described the solution, we now evaluate the regulated on-shell action for the wormhole with fixed cutoff at $u=\epsilon^2$ so that the induced metric on either boundary is 
\begin{equation} \label{bdrymetric}
    ds^2_{\text{bdry}}=\frac{1}{\epsilon^2}\gamma_{ij}dx^i dx^j
\end{equation}
The bulk term evaluates to give
\begin{equation} \label{bulkterm}
    -\frac{1}{16\pi G_N}\int \sqrt{g}(R+d(d-1)-(\nabla \phi)^2)=\frac{d}{8\pi G_N}\int_{\Sigma_d} d^d x\sqrt{\gamma}\int_{\epsilon^2}^{u_0(b)}\frac{du}{u^{\frac{d}{2}+1}\sqrt{1-u+bu^d}}
\end{equation}
where $\Sigma_d$ is the compact $d$-dim Einstein space. Recall that $u_0(b)$ is the smallest positive root of $1-u+bu^d=0$. By evaluating the above integral near the boundary, we see that the divergent pieces are independent of $b$. This observation is crucial to the way we renormalise the action as we shall describe soon. To evaluate the GHY term, note that the extrinsic curvature of the const-$u$ slices is given by
\begin{equation}
    K=d\sqrt{1-u+bu^d}
\end{equation}
So, to leading order as we approach the boundary i.e $u \to 0$, $K=d$. So, the GHY term on each cutoff surface evaluates to give
\begin{equation}
    -\frac{1}{8\pi G}\int \sqrt{h}K=-\frac{d}{8\pi G_N}\frac{\sqrt{1-\epsilon^2+b \epsilon^{2d}}}{\epsilon^d}\int_{\Sigma_d}\sqrt{\gamma}
\end{equation}
We see that in the limit $\epsilon \to 0$, the finite piece of the above GHY term has no dependence on $b$. Note that the divergent cutoff dependent pieces of both the bulk and GHY terms are independent of $b$. Since the induced metric on the boundary \eqref{bdrymetric} is also independent of $b$, the counter-terms needed to render the action finite being functionals of the boundary metric are also indepent of $b$. In summary, these observations show that the difference in the regulated actions of the deformed and undeformed wormholes is finite as $\epsilon \to 0$,
\begin{equation}
    I(b)-I(0)=\lim_{\epsilon \to 0}\frac{d}{8\pi G_N}\int_{\Sigma_d} d^d x\sqrt{\gamma}\,\left (\int_{\epsilon^2}^{u_0(b)}\frac{du}{u^{\frac{d}{2}+1}\sqrt{1-u+bu^d}}-\int_{\epsilon^2}^1 \frac{du}{u^{\frac{d}{2}+1}\sqrt{1-u}}  \right )
\end{equation}
Note that the above difference in actions is also the difference in hyperbolic volumes of these wormholes,
\begin{equation} \label{Janhyp}
    I(b)-I(0)=\frac{d}{8\pi G_N}(V(b)-V(0))
\end{equation}

\subsection{Wormhole volumes and their monotonicity}

We show that the difference in volumes $V(b)-V(0)$ and consequently the difference in actions $I(b)-I(0)$ is positive and increases monotonically with the deformation parameter in $b$. We also derive analytic expressions for these difference. We shall refer to $V_{\text{WH}}(b) \equiv V(b)-V(0)$ as the volume of the deformed wormhole. For even values of $d$, the wormhole volumes are given by
\begin{multline} \label{voleven}
    \frac{V_{\text{WH}}(b)}{V(\Sigma_d)}=b\frac{(d-2)!!}{(d-3)!!}{}_{d}F_{d-1}\bigg [\left (\frac{d+2}{2d},\frac{d+4}{2d},\dots, \frac{2d}{2d},\frac{2d}{2d},\frac{2d+2}{2d},\dots ,\frac{3d-2}{2d}\right ),\\ \left(\frac{d+1}{2(d-1)},\frac{d+3}{2(d-1)},\dots ,\frac{3d-5}{2(d-1)},2 \right ),\frac{b}{b_0}\bigg]
\end{multline}
For odd values of $d$, the wormhole volumes are given by
\begin{multline} \label{volodd}
   \frac{V_{\text{WH}}(b)}{V(\Sigma_d)}=\frac{\pi b}{2}\frac{(d-2)!!}{(d-3)!!}
{}_{d-1}F_{d-2}\bigg [\left (\frac{d+2}{2d},\frac{d+4}{2d},\dots,\frac{3d-2}{2d} \right ),\\ \left (\frac{d+3}{2(d-1)},\frac{d+5}{2(d-1)},\dots ,\frac{3d-5}{2(d-1)},2\right ),\frac{b}{b_0}\bigg ]
\end{multline}
where $V(\Sigma_d)\equiv \int_{\Sigma_d} d^d x \sqrt{\gamma}$ is the hyperbolic volume of the compact $d$-dim hyperbolic space and recall that $b_0=\frac{1}{d}(\frac{d-1}{d})^{d-1}$.

\subsubsection*{Derivation:}

Now, we derive the above formulae. For convenience, we define the function
\begin{equation}
    F(b)\equiv \frac{V_{\text{WH}}(b)}{V(\Sigma_d)}
\end{equation}
We determine $F$ by evaluating all its derivatives at $b=0$ and plugging them into the Taylor series expansion for $F$ around $b=0$,
\begin{equation}
    F(b)=\sum_{n=1}^{\infty}F^{(n)}(0)\frac{b^n}{n!}
\end{equation}
Since $F(0)=0$, we have set the lower limit of the sum to $1$. By evaluating the derivatives $F^{(n)}(0)$, we shall show that this series is absolutely convergent with radius of convergence $b_0$. We evaluate the derivatives using the Leibniz rule. To this end, we write $F$ as a limit,
\begin{equation}
    F(b)=\lim_{\epsilon \to 0}\left (\lim_{\delta \to 0^+}\int_{\epsilon^2}^{u_0(b)-\delta}\frac{du}{u^{\frac{d}{2}+1}\sqrt{1-u+bu^d}}-\int_{\epsilon^2}^1 \frac{du}{u^{\frac{d}{2}+1}\sqrt{1-u}} \right )
\end{equation}
with the regulators $\epsilon$ and $\delta$ chosen to be independent of $b$. Since the integral is convergent in the neighbourhood of $u_0$, the $\delta \to 0$ limit exists. By expanding the integrands near $u=0$, we see that the $\epsilon \to 0$ limit also exists. The two limits on the first integral commute. Applying Leibniz rule, we have
\begin{equation}
    F'(b)=\lim_{\delta \to 0^+}\left (\frac{u_0'}{(u_0-\delta)^{\frac{d}{2}+1}\sqrt{\delta}\sqrt{g(u_0)}}-\frac{1}{2}\int_{0}^{u_0-\delta}\frac{u^{\frac{d}{2}-1} du }{(1-u+bu^d)^{\frac{3}{2}}} \right )
\end{equation}
where we have written $1-u+bu^d=(u_0-u)g(u)$ with $g(u)$ being a $(d-1)$-degree polynomial strictly positive at $u_0$. Since the integral now converges near $u=0$, we have set the lower limit of the integral to $0$. We are interested in the $O(\delta^0)$ terms in the expression inside the limit. Clearly, the first term has no such terms. So, the required term if non-zero must come from the second term in the above expression. We can repeat this procedure to further differentiate to get the higher derivatives as well,
\begin{equation}
    F^{(n)}(b)\overset{O(\delta^0)}\subset\frac{(-1)^n}{2^{n}}(2n-1)!!\int_{0}^{u_0-\delta}\frac{u^{(n-\frac{1}{2})d-1}du}{(1-u+bu^d)^{n+\frac{1}{2}}}
\end{equation}
where $\overset{O(\delta^0)}\subset$ means the LHS is the $O(\delta^0)$ piece of the RHS as $\delta \to 0$.
Now, we can actually evaluate the integral at $b=0$ and extract this $O(\delta^0)$ piece,
\begin{equation}
    F^{(n)}(0)\overset{O(\delta^0)}\subset\frac{(-1)^n}{2^{n}}(2n-1)!!\int_{\delta}^{1}\frac{(1-u)^{(n-\frac{1}{2})d-1}du}{u^{n+\frac{1}{2}}}
\end{equation}
We have thus the expression for the derivatives of $F$ at $b=0$,
\begin{equation}
    F^{(n)}(0)=\frac{(-1)^n }{2^{n}}(2n-1)!!\frac{\Gamma(\frac{1}{2}-n)\Gamma(\frac{1}{2}d(2n-1))}{\Gamma(\frac{1}{2}(d-1)(2n-1))}=\sqrt{\pi}\frac{\Gamma(\frac{1}{2}d(2n-1))}{\Gamma(\frac{1}{2}(d-1)(2n-1))}
\end{equation}
Since the arguments of gamma functions are positive, we see that all the derivatives are positive i.e $F^{(n)}(0)>0$. This proves that the renormalised volumes increase monotonically under the deformation $b$. Thus plugging into the Taylor series expansion for $F$, we have
\begin{equation} \label{Fb}
    F(b)=\sqrt{\pi}\sum_{n=1}^{\infty}\frac{\Gamma(\frac{1}{2}d(2n-1))}{\Gamma(\frac{1}{2}(d-1)(2n-1))}\frac{b^n}{n!}
\end{equation}
The ratio test shows that the radius of convergence of the above series is $b_0=\frac{1}{d}(\frac{d-1}{d})^{d-1}$. This series can be summed to give a closed form expression in terms of hypergeometric functions as recorded in \eqref{voleven} and \eqref{volodd}. For concreteness, let us evaluate $F(b)$ for some special values of $d$. In AdS$_4$, it takes the form
\begin{equation}
    F(b)=\frac{1}{2}\pi b {}_2F_1\left [\frac{5}{6},\frac{7}{6};2;\frac{27b}{4}\right]
\end{equation}
and in AdS$_5$, it takes the form
\begin{equation}
    F(b)=2b {}_4 F_3 \left[\left(\frac{3}{4},1,1,\frac{5}{4} \right),\left (\frac{5}{6},\frac{7}{6},2 \right),\frac{256 b}{27} \right]
\end{equation}

\subsubsection*{Monotonicity:}

We have shown using the above derivation that the wormhole volumes increase monotonically under the deformation $b$. In Appendix \ref{dilatonapp}, we have shown using a similar derivation that the difference in asymptotic values of the dilaton $\Delta \phi$ also increases monotonically with $b$. So, combining the two results, we see that the wormhole volumes increase monotonically with the difference $\Delta \phi$. In fact, we can write down a very simple expression for the rate of increase of the volumes with $\Delta \phi$ in terms of $b$. To see this, note the following nice identity obeyed by the functions $F(b)$ and $G(b)$,
\begin{equation}
    F'(b)=(d-1)(\frac{1}{2}G(b)+bG'(b))
\end{equation}
which can be easily verified using the Taylor series expansions for $F$ and $G$. Recalling the definitions of these functions,
\begin{equation}
  F(b)=\frac{V_{\text{WH}}(b)}{V(\Sigma_d)} \qquad G(b)=\frac{\Delta \phi}{\sqrt{bd(d-1)}}
\end{equation}
we thus have an expression for the rate of volume growth,
\begin{equation} \label{volgrowth}
  \frac{\partial V_{\text{WH}}}{\partial (\Delta \phi)}=\sqrt{\frac{d-1}{d}}\sqrt{b}V(\Sigma_d) >0
\end{equation}
which explicitly confirms monotonicity since RHS is positive. Note that this expression implies that for small deformations $b \ll 1$ or equivalently when the difference in couplings is small, the volume grows quadratically,
\begin{equation}
  V_{\text{WH}} \sim (\Delta \phi)^2 \qquad \Delta \phi \to 0
\end{equation}

\subsection{Wormhole amplitudes between flat boundary metrics}

We now compute the wormhole amplitudes between flat boundary metrics so that we could relate the wormhole amplitudes to averages of two-copy observables computed in flat space. To this end, it is convenient to express the metric on the wormholes discussed above using the Poincare metric on the radial slices,
\begin{equation}
    ds^2=\frac{du^2}{4u^2(1-u+bu^d)}+\frac{1}{uz^2}(dx^i dx_i+dz^2)
\end{equation}
where $x^i$ are the $(d-1)$ transverse coordinates on $H_d$ whose boundary is at $z=0$. The range of the coordinates $(x^i,z)$ is restricted to a fundamental domain of the quotient of $H_d$ by a discrete subgroup of its isometry group.
To compute the on-shell action, we employ a ``wiggly'' cutoff in $u$ near the asymptotic boundary at $u=\frac{\epsilon^2}{z^2}$ so that the induced metric on the cutoff surface is flat to leading order in the cutoff parameter $\epsilon$,
\begin{equation}
  ds^2_{\text{bdry}}=\frac{1}{\epsilon^2}(dx^i dx_i+dz^2)+\frac{dz^2}{z^2}+O(\epsilon^2)dz^2
\end{equation}
Note importantly that the $O(\epsilon^0)$ term in $\sqrt{h}$ does not depend on $b$. By computing the extrinsic curvature of the cutoff surface, it is easy to verify that the $O(\epsilon^0)$ piece of the GHY term does not depend on $b$ so cancels against the GHY term of the undeformed womhole when we subtract the on-shell actions. Now, the bulk term in the action evaluates to
\begin{equation}
   -\frac{1}{16\pi G_N}\int \sqrt{g}(R+d(d-1)-(\nabla \phi)^2)=\frac{d}{8\pi G_N}\int_{\Sigma_d} d^{d-1}x \frac{dz}{z^d}\int_{\frac{\epsilon^2}{z^2}}^{u_0(b)}\frac{du}{u^{\frac{d}{2}+1}\sqrt{1-u+bu^d}}
\end{equation}
We are evaluating the same integral as in \eqref{bulkterm} with a rescaled cutoff which means that the $O(\epsilon^0)$ piece of this integral as $\epsilon \to 0$ remains unchanged. Therefore, the difference in the on-shell actions of the deformed and undeformed wormholes remains unchanged when compared to the calculation done with hyperbolic boundary metrics as expected,
\begin{equation} \label{Janflat}
    I(b)-I(0)=\frac{d}{8\pi G}(V(b)-V(0))=\frac{d}{8\pi G} F(b)V(\Sigma_d)
\end{equation}
 Just like in three dimensions, the Brown-York stress tensor on either boundary remains unaffected by the deformation which means that the deformation is marginal for the two CFTs.

Interpreting the gravity result \eqref{Janhyp} or \eqref{Janflat} in the context of the famous example of the AdS$_5$/CFT$_4$ duality \cite{Maldacena:1997re}, this gives a concrete prediction for the correlation between the partition functions of two copies of $\mathcal{N}=4$ SYM on a compact hyperbolic space $\Sigma_4$ and deformed by different couplings to a marginal scalar dual to the dilaton at large-$N$,
\begin{equation}
    \overline{Z_{g^2_{YM+}}[\Sigma_4]Z_{g^2_{YM-}}[\Sigma_4]}=\exp\left (-\frac{N^2}{\pi^2}F(b)V(\Sigma_4)\right )\overline{Z_{g^2_{YM}}[\Sigma_4]Z_{g^2_{YM}}[\Sigma_4]}
\end{equation}
where $V(\Sigma_4)\equiv \int_{\Sigma_4}d^4x \sqrt{\gamma}$ is the hyperbolic volume of $\Sigma_4$. The relation between the bulk and boundary parameters is given by
\begin{equation}
    G_5=\frac{\pi}{2N^2},\qquad e^{\phi_0}=\frac{g_{YM}^2}{2\pi},\qquad e^{\phi_0\pm \frac{\Delta \phi(b)}{2}}=\frac{g^2_{YM\pm}}{2\pi}
\end{equation}
where $G_5$ is the 5d Newton's constant and $\Delta \phi(b)$ is given by \eqref{ads5deltaphi}. For small differences $\Delta \phi$ in which case the ratio of couplings $\frac{g_{YM+}^2}{g_{YM-}^2}$ is close to $1$, the above decay of correlation can be expressed as
\begin{equation} \label{corrsmall}
   \frac{\overline{Z_{g^2_{YM+}}[\Sigma_4]Z_{g^2_{YM-}}[\Sigma_4]}}{\overline{Z_{g^2_{YM}}[\Sigma_4]Z_{g^2_{YM}}[\Sigma_4]}} \approx \exp \left (-\frac{3N^2}{32\pi^2}\left (\frac{g_{YM+}^2}{g_{YM-}^2}-1 \right )^2 V(\Sigma_4) \right )
\end{equation}

\subsection{Dilaton correlators from wormhole amplitudes} \label{dilhigherd}

Just like in the three dimensional case, we can compute the integrated dilaton correlators from wormhole amplitudes treating the wormhole amplitude as a generating functional,
\begin{equation}
    Z_{\text{WH}}[\phi_+,\phi_-;\Sigma_d] = \overline{\left \langle \exp \left(-\phi_+\int_{\Sigma_d} d^d x \sqrt{g}  \mathcal{O}(x)\right) \right \rangle \left \langle \exp \left(-\phi_-\int_{\Sigma_d} d^d x \sqrt{g}  \mathcal{O}(x)\right) \right \rangle}
\end{equation}
where $\Sigma_d$ is the compact hyperbolic space with volume $V(\Sigma_d)$. The operator $\mathcal{O}$ is the marginal scalar (has scaling dimension $\Delta=d$) dual to the dilaton. The asymptotic values $\phi_\pm$ of the dilaton are sources for $\mathcal{O}$ in the CFT path integral. The integrals are over a fundamental domain of the quotient of $H_d$ that corresponds to $\Sigma_d$.
The wormhole amplitude rescaled by the partition function of the undeformed wormhole was computed to be
\begin{equation}
   Z_{\text{WH}}=\exp\left (-\frac{d}{8\pi G_N}F(b)V(\Sigma_d) \right )
\end{equation}
with $F(b)$ given in \eqref{Fb}.
The integrated one-point functions $G_1^\pm$ on either boundary defined as
\begin{equation}
  G_1^\pm= \frac{\overline{ \left \langle \int_{\Sigma_{d}} d^d x  \sqrt{g}    \mathcal{O}(x) \exp \left(-\phi_\pm\int_{\Sigma_{d}} d^d x  \sqrt{g}  \mathcal{O}(x)\right) \right \rangle   \left \langle  \exp \left(-\phi_\mp\int_{\Sigma_{d}} d^d x  \sqrt{g}  \mathcal{O}(x)\right) \right \rangle }}{\overline{\left \langle \exp \left(-\phi_+\int_{\Sigma_{d}} d^d x  \sqrt{g}  \mathcal{O}(x)\right) \right \rangle \left \langle \exp \left(-\phi_-\int_{\Sigma_{d}} d^d x  \sqrt{g} \mathcal{O}(x)\right) \right \rangle}}
\end{equation}
 are given by
\begin{equation}
 G_1^\pm =-\frac{\partial \log Z_{\text{WH}}}{\partial  \phi_\pm}=\pm\frac{\sqrt{d(d-1)}}{8\pi G_N}V(\Sigma_d)\sqrt{b}
\end{equation}
where we used the fact that upto proportionality factors, the integrated one-point function is given by rate of volume growth computed in \eqref{volgrowth}.

Now, we turn to the integrated two-point functions.
The connected self-correlators $G_2^\pm$ defined by the integrated two-point functions of the marginal scalar on the same boundary, 
\begin{align}
  & G^\pm_2=-(G_1^\pm)^2+\notag \\ & \frac{\overline{ \left \langle \int_{\Sigma_{d}} d^d x d^d y   \sqrt{g(x)g(y)}   \mathcal{O}(x) \mathcal{O}(y)\exp \left(-\phi_\pm\int_{\Sigma_{d}} d^d x   \sqrt{g} \mathcal{O}(x)\right) \right \rangle   \left \langle  \exp \left(-\phi_\mp\int_{\Sigma_{d}} d^d x  \sqrt{g}   \mathcal{O}(x)\right) \right \rangle }}{\overline{\left \langle \exp \left(-\phi_+\int_{\Sigma_{d}} d^d x  \sqrt{g}  \mathcal{O}(x)\right) \right \rangle \left \langle \exp \left(-\phi_-\int_{\Sigma_{d}} d^d x  \sqrt{g} \mathcal{O}(x)\right) \right \rangle}}
\end{align}
and the connected cross-correlators $G_{1,1}$ defined by the integrated two-point functions of the marginal scalar on different boundaries,
\begin{align}
& G_{1,1}=-G_1^+ G_1^-+\notag \\ & \frac{\overline{ \left \langle \int_{\Sigma_{d}} d^d x \sqrt{g}  \mathcal{O}(x)\exp \left(-\phi_+\int_{\Sigma_{d}} d^d x   \sqrt{g}  \mathcal{O}(x)\right) \right \rangle   \left \langle  \int_{\Sigma_d} d^d y  \sqrt{g}   \mathcal{O}(y)\exp \left(-\phi_-\int_{\Sigma_{d}} d^d x  \sqrt{g}  \mathcal{O}(x)\right) \right \rangle }}{\overline{\left \langle \exp \left(-\phi_+\int_{\Sigma_{d}} d^d x  \sqrt{g}  \mathcal{O}(x)\right) \right \rangle \left \langle \exp \left(-\phi_-\int_{\Sigma_{d}} d^d x \sqrt{g}  \mathcal{O}(x)\right) \right \rangle}}
\end{align}
are given by
\begin{equation}
\begin{split}
   G_2^\pm =&  \frac{\partial ^2 \log Z_{\text{WH}}}{\partial \phi_\pm^2}= -\frac{1}{8\pi G_N}\frac{V(\Sigma_d)}{G(b)+2bG'(b)}\\
     G_{1,1} = &  \frac{\partial ^2 \log Z_{\text{WH}}}{\partial \phi_+\partial \phi_-} =  \frac{1}{8\pi G_N}\frac{V(\Sigma_d)}{G(b)+2bG'(b)}
 \end{split} 
\end{equation}
where we used $\Delta \phi =\sqrt{bd(d-1)} G(b)$. Just like in three dimensions, we see that the self-correlators are negative and cross-correlators are positive. Using the monotonicity of the function $G(b)$ we see that $G(b)+2bG'(b) \geq G(0)$. So, the magnitude of the correlators decrease monotonically under the deformation from the maximum value,
\begin{equation} \label{twopointhigherd}
   G_{1,1}=|G_2^\pm| \leq \frac{V(\Sigma_d)}{8\sqrt{\pi^3}G_N}\frac{\Gamma(\frac{d+1}{2})}{\Gamma(\frac{d}{2})}
\end{equation}
where we used that $G(0)=\sqrt{\pi}\frac{\Gamma(\frac{d}{2})}{\Gamma(\frac{d+1}{2})}$. The decay of the cross-correlator further illustrates the decay in correlation between the CFTs as the difference in marginal couplings grows. For $d=2$, the results in this section are consistent with those derived in section \ref{dil3d}.

\section{Discussion} \label{discussion}

In this paper, we constructed a one-parameter family of Euclidean wormhole solutions in Einstein-dilaton gravity in general dimensions which are locally the same as the single-boundary Janus domain wall solutions. They compute the averaged product of partition functions of CFTs on either boundary deformed by different marginal couplings to the scalar dual to the dilaton. We observed that the renormalised volumes of these wormholes increase monotonically with the difference in the marginal couplings on the boundary thereby showing that the pair of CFTs on the boundaries get increasingly decorrelated as the difference in the marginal couplings increases. We used the partition functions of the three-dimensional wormhole solutions to determine the variance of the OPE data between the marginally deformed 2d CFTs and quantified how the variance decays with the difference in marginal couplings. Treating the wormhole amplitudes as generating functionals for dilaton correlators, we computed various integrated dilaton correlators and in particular, we observed that the crossed two-point correlators with a dilaton insertion on either boundary decay montonically with the difference in marginal couplings consistent with our observation that the CFTs increasingly decorrelate as the difference in marginal couplings grows.

Now, we discuss some interesting directions closely related to this paper which we hope to explore in future works:

\subsection{Averaging over marginal couplings}

In this work, we described a one-dimensional space of wormhole deformations. We observed that the integrated dilaton two-point functions are positive and bounded in this deformation space. This makes these two-point functions as natural candidates for a choice of metric on this deformation space. This is analogous to the Zamolodchikov metric on the moduli space of exactly marginal deformations of CFTs which is given by the two-point functions of the marginal operators. It would be interesting to explore the consequences of making the deformation $\Delta \phi$ random and averaging over the deformation space using this metric given by the integrated dilaton two-point function. For instance, using the integrated two point function $G_{1,1}(\Delta \phi)$ computed in \eqref{twopoint3d}, we define `quenched' and `annealed' averages of wormhole amplitudes respectively as $\langle \log Z_{\text{WH}} \rangle$ and $\log \langle Z_{\text{WH}} \rangle$,
\begin{equation}
   \begin{split}
       \langle \log Z_{\text{WH}} \rangle \equiv & \frac{\int_0^{\infty} d(\Delta \phi) G_{1,1}(\Delta \phi) \log Z_{\text{WH}}}{\int_0^{\infty} d(\Delta \phi) G_{1,1}(\Delta \phi)}\\
         \langle  Z_{\text{WH}} \rangle \equiv & \frac{\int_0^{\infty} d(\Delta \phi) G_{1,1}(\Delta \phi)  Z_{\text{WH}}}{\int_0^{\infty} d(\Delta \phi) G_{1,1}(\Delta \phi)} 
   \end{split}
\end{equation}
 In 3d, using $G_{1,1}(\Delta \phi)=\frac{c|\chi|}{12}\text{sech}^2(\Delta \phi)$ and $Z_{\text{WH}}=(\cosh(\Delta \phi))^{\frac{c\chi}{6}}$, the quenched average evaluates to 
\begin{equation}
  \langle \log Z_{\text{WH}} \rangle=\frac{c\chi}{6}\int_0^{\infty} d(\Delta \phi) \text{sech}^2(\Delta \phi)\log (\cosh(\Delta \phi))=-\frac{c|\chi|}{6}(1-\log 2)
\end{equation}
while the annealed average evaluates to
\begin{equation}
  \log \langle  Z_{\text{WH}} \rangle =\log \left( \int_0^{\infty} d(\Delta \phi) \text{sech}^2(\Delta \phi) (\cosh(\Delta \phi))^{\frac{c\chi}{6}} \right) \to -\frac{1}{2}\log \left (\frac{c|\chi|}{3\pi} \right )
 \end{equation}
where upon evaluating the integral, we have only retained the leading term at large-$c$ in the expression for the annealed average. Recall that be definition, $Z_{\text{WH}}$ is the partition function on the deformed wormhole rescaled by the partition function on the undeformed wormhole. We can analogously define such averages for the higher dimensional wormhole amplitudes using the integrated two-point function computed in \eqref{twopointhigherd}. The idea of averaging over moduli space of marginal deformations discussed in this section has already been explored in the context of compact free boson CFTs in $2d$ where the averaging was done over the Narain moduli space \cite{Maloney:2020nni,Afkhami-Jeddi:2020ezh}. However, the bulk theory in that case was not Einstein gravity but rather an exotic theory of gravity termed ``U(1)-gravity''.

\subsection{A family of closed universes in three dimensions}

We describe a model of a closed universe by a suitable analytic continuation of the Janus-deformed wormhole solutions of 3d gravity to Lorentzian signature. The metric on the cosmology is given by analytic continuation $\rho \to it$,
\begin{equation}
    ds^2=-dt^2+\frac{1}{2}(1+a^2 \cos(2t))d\Sigma^2
\end{equation}
where $d\Sigma^2$ is the hyperbolic metric on the compact Riemann surface $\Sigma$. Note that to get a real solution for the dilaton, it is necessary that $a \geq 1$ which is to be contrasted with the regime $a \leq 1$ where the Euclidean wormhole solutions are constructed. This can be seen by looking at the $tt$ component of Einstein's equations,
\begin{equation}
    R_{\mu \nu}+2 g_{\mu \nu}=\partial_\mu \phi \partial_\nu \phi
\end{equation}
The $tt$ component reads
\begin{equation}
    R_{tt}+2 g_{tt}=\frac{2(a^4-1)}{(1+a^2 \cos(2t))^2}=\Dot{\phi}^2
\end{equation}
It is interesting to note that if $a<1$, the solutions do not have the cosmological singularities. So, naively it might appear that turning on the Janus-like deformation can resolve the cosmological singularities. However, note that such a solution is sourced by a scalar field with negative kinetic energy which violates energy conditions. 

In the subsequent discussion we restrict to the regime of $a > 1$. The $a=1$ cosmologies have been discussed in \cite{Maldacena:2004rf}. The dilaton profile $\phi=\phi(t)$ is given by
\begin{equation}
    \phi(t)=\phi_0+\sqrt{2}\tan^{-1}\left (\sqrt{\frac{a^2-1}{a^2+1}}\tan(t) \right )
\end{equation}
In the limit $a \to \infty$, we see that the dilaton increases linearly with proper time $\phi=\phi_0+\sqrt{2}t$.
A consequence of the condition $a > 1$ is that the metric above describes a bang-crunch cosmology with the crunch and bang occurring respectively at
\begin{equation}
    \tan(t_\pm)=\pm\sqrt{\frac{a^2+1}{a^2-1}} \implies \Delta t=2\tan^{-1}\left (\sqrt{\frac{a^2+1}{a^2-1}} \right )
\end{equation}
where $\Delta t$ is the proper time between the bang and crunch and takes values in the range $(\frac{\pi}{2},\pi)$ as we vary $a$ with the lower bound occurring at $a \to \infty$ and the upper bound at $a \to 1$. 
When the deformation parameter is large,
\begin{equation}
    t_\pm \to \pm \frac{\pi}{4} \qquad a \to \infty
\end{equation}
For each value of $a$, the dilaton increases monotonically in the interval $\Delta t$ so could be a potential candidate for a physical clock in these closed universes.
At the bang and crunch, the dilaton takes values
\begin{equation}
    \phi(t_\pm)=\phi_0 \pm \frac{\pi}{2\sqrt{2}} \implies \Delta\phi =\frac{\pi}{\sqrt{2}}
\end{equation}
independent of the value of $a$. So, if the dilaton is used as a clock in this universe, then the time elapsed on the clock between the bang and crunch is a constant and independent of $a$. The cosmological models described in this section are 3d analogues of models of closed universes in JT gravity constructed in \cite{Usatyuk:2024mzs} by analytically continuing wormhole solutions in JT gravity. See also \cite{Antonini:2023hdh} for models of cosmologies in AdS$_{d+1}$ sourced by thin shells of dust obtained by analytically continuing single-boundary Euclidean geometries sourced by the dust shells. It would be interesting to understand the boundary interpretation of such closed universes.

\subsection{CFT ensembles in higher dimensions}

In section \ref{higherdwormholes}, we derived the correlation between the partition functions of marginally deformed CFTs in general dimenions by a gravity calculation,
\begin{equation}
  \overline{Z[\Sigma_d]^{\phi_+} Z[\Sigma_d]^{\phi_-}}=\exp\left(- \frac{d}{8\pi G_N}F(b)V(\Sigma_d)\right) \overline{Z[\Sigma_d] Z[\Sigma_d]}
\end{equation}
It would be interesting to explicitly derive the average between the partition functions of the undeformed theories $\overline{Z[\Sigma_d] Z[\Sigma_d]}$ in the above expression by calculating the renormalised on-shell action of the undeformed wormhole, and understand the implication in terms of possible averaging over OPE data of CFT$_d$ analogous to the story in three dimensions \cite{Chandra:2022bqq}. Specifically, it would be nice if such a wormhole computation encodes the universal asymptotics of the OPE data of global conformal primaries in higher dimensional CFTs computed in \cite{Benjamin:2023qsc}.

\subsection{Dilaton-matter couplings}

In sections \ref{OPE} and \ref{Linedef} where we added matter worldlines to Einstein-dilaton gravity in three dimensions, we assumed that the dilaton doesn't couple to the matter worldlines. But it would be interesting to understand the effects of explicit dilaton coupling to the worldlines for example described by the term in the action,
\begin{equation}
  I_{\text{matter}}=-m \int \phi ds
\end{equation}
on the OPE statistics of the dual operators and to understand if such sub-threshold operators gain anomalous dimensions due to the explicit coupling to the dilaton.

\ \\ 

\ \\

\noindent \textbf{Acknowledgments}\\

I would like to specially thank Tom Hartman for very useful discussions on this work and for collaboration on related projects. I also thank Viraj Meruliya for collaboration on related projects. I thank Alex Maloney for explaining an ongoing project with Clement Virelly on related topics. This work was completed during my visit to Tata Institute of Fundamental Research (TIFR), India. I thank the string theory group at TIFR and especially Shiraz Minwalla for the hospitality and stimulating discussions about ensemble averaging.
This work was supported by NSF grant PHY-2014071.

\appendix

%\addtocontents{toc}{\setcounter{tocdepth}{0}}

\section{Asymptotic values of the dilaton in general dimensions} \label{dilatonapp}

By integrating the equation of motion for the dilaton, we want to evaluate the difference in the asymptotic values of the dilaton denoted as $\Delta \phi$,
\begin{equation}
    \Delta \phi= c \int_{0}^{u_0}\frac{u^{\frac{d}{2}-1}du}{\sqrt{1-u+bu^d}}
\end{equation}
with the integration constant $c$ fixed by Einstein's equations to give
\begin{equation}
  c=\sqrt{b d(d-1)}
\end{equation}
Clearly, $c$ is not analytic in $b$ at $b=0$ which means $\Delta \phi$ treated as a function of $b$ is also not analytic at $b=0$. 
To this end, we define the function,
\begin{equation}
    G(b) \equiv \frac{\Delta \phi}{c}=\int_{0}^{u_0(b)}\frac{u^{\frac{d}{2}-1}du}{\sqrt{1-u+bu^d}}
\end{equation}
which is analytic at $b=0$.
Now, we evaluate $G(b)$ by computing all its derivatives $G^{(n)}(0)$ and plugging into its Taylor series expansion around $b=0$,
\begin{equation}
    G(b)=\sum_{n=0}^{\infty}G^{(n)}(0)\frac{b^n}{n!}
\end{equation}
In order to apply Leibniz rule, we introduce a regulator $\delta$ near $u_0$ and express $G(b)$ as a limit,
\begin{equation}
    G(b)=\lim_{\delta \to 0^+}\int_0^{u_0(b)-\delta}\frac{u^{\frac{d}{2}-1}du}{\sqrt{1-u+bu^d}}
\end{equation}
which is well-defined.
Applying Leibniz rule, we have
\begin{equation}
    G'(b)=\lim_{\delta \to 0^+}\left (\frac{u_0'(u_0-\delta)^{\frac{d}{2}-1}}{\sqrt{\delta}\sqrt{g(u_0)}}-\frac{1}{2}\int_{0}^{u_0-\delta}\frac{u^{\frac{3d}{2}-1} du }{(1-u+bu^d)^{\frac{3}{2}}} \right )
\end{equation}
where we have written $1-u+bu^d=(u_0-u)g(u)$ with $g(u)$ being a $(d-1)$-degree polynomial strictly positive at $u_0$. We are interested in the $O(\delta^0)$ terms in the expression inside the limit. Clearly, the first term has no such terms so does not survive the limit. So, the required term if non-zero must come from the second term in the above expression. We can repeat this procedure to further differentiate to get the higher derivatives as well,
\begin{equation}
    G^{(n)}(b)\overset{O(\delta^0)}\subset\frac{(-1)^n}{2^{n}}(2n-1)!!\int_{0}^{u_0-\delta}\frac{u^{(n+\frac{1}{2})d-1}du}{(1-u+bu^d)^{n+\frac{1}{2}}}
\end{equation}
where $\overset{O(\delta^0)}\subset$ means the LHS is the $O(\delta^0)$ piece of the RHS as $\delta \to 0$.
Now, we can actually evaluate the integral at $b=0$ and extract this $O(\delta^0)$ piece,
\begin{equation}
    G^{(n)}(0)\overset{O(\delta^0)}\subset\frac{(-1)^n}{2^{n}}(2n-1)!!\int_{\delta}^{1}\frac{(1-u)^{(n+\frac{1}{2})d-1}du}{u^{n+\frac{1}{2}}}
\end{equation}
The $O(\delta^0)$ term in the integral is given by
\begin{equation}
   \int_{\delta}^{1}\frac{(1-u)^{(n+\frac{1}{2})d-1}du}{u^{n+\frac{1}{2}}}\overset{O(\delta^0)}\supset \frac{\Gamma(\frac{1}{2}-n)\Gamma(\frac{1}{2}d(2n+1))}{\Gamma(n(d-1)+\frac{d+1}{2})}
\end{equation}
We have thus the expression for the derivatives of $F$ at $b=0$,
\begin{equation}
    G^{(n)}(0)=\sqrt{\pi}\frac{\Gamma(\frac{1}{2}d(2n+1))}{\Gamma(n(d-1)+\frac{d+1}{2})}
\end{equation}
where we used the identity
\begin{equation}
  \frac{(-1)^n}{2^n}(2n-1)!! \Gamma(\frac{1}{2}-n)=\sqrt{\pi}
\end{equation}
Since the arguments of the gamma functions are positive, we see that all the derivatives are positive i.e $G^{(n)}(0)>0$. The expression as written above also holds for $n=0$ i.e gives $ G(0)=\sqrt{\pi}\frac{\Gamma(\frac{d}{2})}{\Gamma(\frac{d+1}{2})}$
which is also positive. Upto positive prefactors, since $\Delta \phi \sim \sqrt{b}G(b)$, this shows that $\Delta \phi$ increases monotonically with $b$. Plugging into the Taylor series expansion for $G$, we have
\begin{equation}
   \frac{ \Delta \phi}{c}=G(b)=\sqrt{\pi}\sum_{n=0}^{\infty}\frac{\Gamma(\frac{1}{2}d(2n+1))}{\Gamma(n(d-1)+\frac{d+1}{2})}\frac{b^n}{n!}
\end{equation}
The ratio test shows that the radius of convergence of the above series is $b_0=\frac{1}{d}(\frac{d-1}{d})^{d-1}$. This series can be summed to give a closed form expression in terms of generalized hypergeometric functions, 
\begin{multline}
    \Delta \phi =a_d\sqrt{bd(d-1)}\frac{(d-2)!!}{(d-1)!!} {}_d F_{d-1}\bigg [\left ( \frac{d}{2d},\frac{d+2}{2d},\dots, \frac{3d-2}{2d}\right),\\ \left ( \frac{d+1}{2(d-1)},\frac{d+3}{2(d-1)},\dots, \frac{3d-5}{2(d-1)},\frac{3}{2}\right),\frac{b}{b_0(d)}\bigg]
\end{multline}
where the coefficient $a_d$ takes different values for even and odd $d$,
\begin{equation}
    a_d=
    \begin{cases}
         2 \qquad & d \in 2\mathbb{Z}_+ \\
         \pi \qquad & d \in 2\mathbb{Z}_+ +1
    \end{cases}
\end{equation}
Although we have computed $\Delta \phi$ to all orders in $b$, for applications where $b$ is small it is useful to explicitly write out the first few terms in the series,
\begin{equation}
  \Delta \phi =\sqrt{\pi d(d-1)}\left(\frac{\Gamma(\frac{d}{2})}{\Gamma(\frac{d+1}{2})}b^{\frac{1}{2}}+\frac{\Gamma(\frac{3d}{2})}{\Gamma(\frac{3d-1}{2})}b^{\frac{3}{2}}+O(b^{\frac{5}{2}})\right )
\end{equation}

\section{Review: Liouville solutions for thin shell wormholes in 3d gravity} \label{thinshellapp}

We review the Liouville solutions that correspond to two-boundary wormhole solutions in 3d gravity sourced by thin shells of dust particles closely following \cite{Chandra:2024vhm}. In particular, we focus on two such wormhole solutions: one which only has a single thin shell going across and threreby computes holographically the variance of the $g$-function of the dual line defect \eqref{linedef},
\begin{equation}
   Z_{\text{grav}}=\overline{|\langle D_\Sigma \rangle |^2}
\end{equation}
and the other which has the thin shell and the worldlines of two massive point particles going across and thereby computes holographically the variance of the matrix elements of the dual line defect \eqref{linedef},
\begin{equation}
  Z_{\text{grav}}=\overline{|\bra{i} D_\Sigma \ket{j}|^2}
\end{equation}
The metric on these wormholes can be expressed in hyperbolic slicing as
\begin{equation}
  ds^2=d\rho^2+\cosh^2(\rho)e^{\Phi}|dz|^2
\end{equation}
with the corresponding Liouville solutions for $\Phi$ constructed subsequently. The on-shell gravitational actions for these wormholes after suitable renormalisation match with two copies of the corresponding Liouville actions \cite{Chandra:2024vhm}. This means the variances of the matrix elements of $D_\Sigma$ are governed by the Liouville actions. So, we now discuss these Liouville solutions.

\subsection{Liouville solution with one line defect on the sphere}

We describe the construction of the Liouville solution that calculates the 1-point function of the semiclassical line defect,
\begin{equation} \label{quantumL}
   L_\Sigma = \exp\left[{\frac{m}{4\pi} \int_\Sigma d\ell \Phi } \right]
\end{equation}
placed along a closed curve $\Sigma$ on the sphere in Liouville CFT.
 It is convenient to work in the cylinder conformal frame with $z \sim z+2\pi R$, and for simplicity, we place the defect of radius $R$ and mass $m$ on the circle Im $z = 0$. 
Provided $mR > 2$, there is a classical saddlepoint in which case the classical solution $e^{\Phi}|dz|^2$ is constructed from two hyperbolic disks glued on a circular interface. 
Thus the Liouville solution used to calculate $\langle L_\Sigma\rangle$ is:
\begin{align}\label{phi1pt}
\Phi = \begin{cases}
-2\log \left (R \sinh \left(A+ \frac{y}{R}\right)\right ) & y > 0\\
-2\log \left (R\sinh \left( A- \frac{y}{R} \right)\right ) & y< 0
\end{cases}
\end{align}
with $z = x + i y$, and $A>0$. With the chosen ansatz for the Liouville solution, the Liouville field is already continuous across the defect. The junction condition for the Liouville field reads 
\begin{equation}
   \partial_y \Phi|_{+} -\partial_y \Phi |_{-} =-2m
\end{equation}
where $+$ is the limit from $y>0$ and $-$ is the limit from $y<0$.
Solving the junction determines the constant $A$ in the above Liouville solution,
\begin{align}
A = \frac{1}{2}\log \frac{mR+2}{mR-2} \ .
\end{align}
%The geometry $e^{\Phi}|dz|^2$ is diffeomorphic to two hyperbolic disks, glued on a circular interface. 
The stress tensor associated with the Liouville solution \eqref{phi1pt} is 
\begin{equation}
  T^{\Phi}(z)=\frac{1}{2}\partial^2\Phi-\frac{1}{4}(\partial \Phi)^2=\frac{1}{4R^2}+\frac{m}{2R}\delta(y)
\end{equation}
so the presence of the line defect shows up as a $\delta$-function in the stress tensor, with the mass parameter $m$ determining the energy density on the defect.
 The regulated action, including a single defect $L_\Sigma$ is
\begin{align}\label{cyl1paction}
S_L = \frac{1}{4\pi}\int_{\Gamma} d^2z\left( \p \Phi \bar{\p}\Phi +e^{\Phi}\right) - \frac{m}{4\pi} \int_\Sigma dz \Phi+ \frac{1}{4\pi R}\int_{\Gamma_+\sqcup \Gamma_-} \!\! dz \Phi + \frac{T}{R}+2(1-\log2)
\end{align}
where $\Gamma$ is the region $y \in [-T,T]$ and $\Gamma_{\pm}$ are the boundaries at $y = \pm T$. 
Evaluating the action \eqref{cyl1paction} on the solution \eqref{phi1pt} we find 
\begin{align} \label{onshell}
S_L &= mR(\log 2+1) - \frac{1}{2}(mR+2)\log(mR+2) - \frac{1}{2}(mR-2)\log(mR-2) \ .
\end{align}
Thus the one-point function of the line defect \eqref{quantumL} is
\begin{align}\label{defect1}
\langle L_\Sigma\rangle \approx \exp\left(-\frac{c}{6}S_L\right)= \left((2e)^{-mR}(mR-2)^{\frac{mR}{2}-1}(mR+2)^{\frac{mR}{2}+1}\right)^{\frac{c}{6}} 
\end{align}
It is convenient to scale the radius to $1$. The hyperbolic area of the sphere with one line defect is given by
\begin{equation}
   \int d^2 z e^{\Phi}=2\pi(m-2) \implies \chi \equiv 2-m
\end{equation}
and we have defined an analogue of the Euler characteristic for the sphere with a line defect in terms of its hyperbolic area. Since the saddle exists only for $m>2$, we see that $\chi<0$ just like the case for smooth hyperbolic surfaces. The one-point function computed in \eqref{defect1} governs the variance of the $g$-function of the CFT defect $D_\Sigma$ with the relation between them given by
\begin{equation}
 \overline{|D_\Sigma|^2}=\mathcal{N}^2 \langle L_\Sigma \rangle^2\qquad \mathcal{N}=e^{-\frac{c}{6}m\log 2}
\end{equation}
where the proportionality factor $\mathcal{N}$ arises from a relative normalisation between the probe local operators in CFT constituting $D_\Sigma$ and the corresponding Liouville vertex operators which constitute $L_\Sigma$ and was derived in \cite{Chandra:2024vhm}.

\subsubsection*{Planar limit}

In the planar limit where $mR \gg 1$, the above expressions simplify considerably. The Liouville solution on the plane with the defect placed along the real axis is given by
\begin{align}
\Phi = \begin{cases}
-2\log (y+\frac{2}{m} ) & y > 0\\
-2\log (\frac{2}{m}-y) & y< 0
\end{cases}
\end{align}
$m$ is now a dimensionless parameter.
Notice that in the limit $m \to \infty$, the above Liouville solution becomes singular on the defect.
The action does not require a regulator in the $y$-direction and no boundary terms are needed at $y \to \pm \infty$,
\begin{align}
S_L = \frac{1}{4\pi}\int_{\Gamma} d^2z\left( \p \Phi \bar{\p}\Phi +e^{\Phi}\right) - \frac{m}{4\pi} \int_\Sigma dz \Phi
\end{align}
The action density (action per unit length) denoted $\Tilde{S}_L$ is uniform along the length of the defect and evaluates to give
\begin{equation}
  \Tilde{S}_L=\frac{m}{2\pi}\log(\frac{2e}{m})
\end{equation}
which matches with the planar limit of \eqref{onshell}. In the planar limit of the sphere with a line defect, the analogue of Euler characteristic defined above becomes $ \chi=-m$.

\subsection{Liouville solution with one line defect and two conical defects}

Now, we discuss the Liouville solution used to compute the variance of the matrix element, $\overline{|\bra{i} D_\Sigma \ket{j}|^2}$ which upto normalisation is governed by the matrix element of the Liouville defect, $\bra{i}L_\Sigma\ket{j}$ \cite{Chandra:2024vhm}. Working in the cylinder frame with the radius of the cylinder rescaled to $1$, the Liouville solution used to compute the resulting matrix element between conical defect states of conformal weights $h=\frac{c}{6}\eta_i(1-\eta_i)$ and $h'=\frac{c}{6}\eta_j(1-\eta_j)$ is:
\begin{align}\label{phi1ptdef}
\Phi = \begin{cases}
-2\log \left( \frac{1}{1-2\eta_i}\sinh\left((1-2\eta_i)(y+ A)\right)\right) & y > 0\\
-2\log \left( \frac{1}{1-2\eta_j}\sinh\left((1-2\eta_j)(A'-y)\right)\right)  & y< 0
\end{cases}
\end{align}
$A$ and $A'$ are determined by solving the continuity and junction conditions for the Liouville field across the line defect at $y=0$. One can use the solution to compute the on-shell action as described in \cite{Chandra:2024vhm}. For the purpose of this work, we just want to emphasize that the hyperbolic area of this surface is given by
\begin{equation}
   \int d^2z e^{\Phi}=2\pi(2\eta_i+2\eta_j+m-2) \implies \chi\equiv 2-m-2\eta_i-2\eta_j
\end{equation}
using which we have defined the generalized Euler characteristic for this surface. So, the saddle exists provided $\chi <0$. 

\subsection{Liouville solution on the torus with two line defects} \label{Liouvtor}

We now review the essential features of the Liouville solution that computes the two-point function $\langle L(\tau_0)L(-\tau_0)\rangle_{\beta}$ on the torus. This Liouville correlator computes the variance of the thermal two-point function of the CFT defect $D_\Sigma$ which was used in section \ref{Linedef}. The torus is described by the following identifications defects at $\pm \tau_0$, but now
\begin{align}
z \sim z + 2\pi \sim z + i \beta \ . 
\end{align}
The classical solution $e^{\Phi}|dz|^2$ consists of two finite-length hyperbolic cylinders, glued at their ends. Hence the Liouville field is
\begin{align}
\Phi = \begin{cases}
-2 \log \left( \frac{1}{r_H} \cos(r_H y)\right) & |y| < \tau_0 \\
-2\log\left( \frac{1}{r_H'}\cos(r_H' (\frac{\beta}{2}-|y|))\right) &  \tau_0 < |y| < \frac{\beta}{2} \ .
\end{cases}
\end{align}
Continuity and the junction condition determine $(r_H, r_H')$ in terms of $(\tau_0, \beta)$. The important relation for the present purpose is
\begin{align}
 \sqrt{e^{\Phi_0}-r_H^2}+\sqrt{e^{\Phi_0}-r_H'^2}=m
\end{align}
where $\Phi_0$ is the Liouville field at the location of either of the defects $y=\pm \tau_0$.
Using this relation, we see that the hyperbolic area of the surface is given by
\begin{equation}
  \int d^2 z e^\Phi = 4\pi m
\end{equation}
thereby proving the result used in \eqref{torchi}.

%\small
\bibliographystyle{ourbst}
\bibliography{biblio1.bib}

\providecommand{\href}[2]{#2}\begingroup\raggedright\begin{thebibliography}{10}

\bibitem{Maldacena:2004rf}
J.~M. Maldacena and L.~Maoz, {{Wormholes in AdS}},
  \href{http://dx.doi.org/10.1088/1126-6708/2004/02/053}{JHEP {\bf 02}, 053,
  2004},
  [\href{http://arxiv.org/abs/arXiv:hep-th/0401024}{{arXiv:hep-th/0401024}}].

\bibitem{Witten:1999xp}
E.~Witten and S.-T. Yau, {{Connectedness of the boundary in the AdS / CFT
  correspondence}}, \href{http://dx.doi.org/10.4310/ATMP.1999.v3.n6.a1}{Adv.
  Theor. Math. Phys. {\bf 3}, 1635--1655, 1999},
  [\href{http://arxiv.org/abs/arXiv:hep-th/9910245}{{arXiv:hep-th/9910245}}].

\bibitem{Cotler:2020ugk}
J.~Cotler and K.~Jensen, {{AdS$_{3}$ gravity and random CFT}},
  \href{http://dx.doi.org/10.1007/JHEP04(2021)033}{JHEP {\bf 04}, 033, 2021},
  [\href{http://arxiv.org/abs/arXiv:2006.08648}{{arXiv:2006.08648 [hep-th]}}].

\bibitem{DiUbaldo:2023qli}
G.~Di~Ubaldo and E.~Perlmutter, {{AdS$_{3}$/RMT$_{2}$ duality}},
  \href{http://dx.doi.org/10.1007/JHEP12(2023)179}{JHEP {\bf 12}, 179, 2023},
  [\href{http://arxiv.org/abs/arXiv:2307.03707}{{arXiv:2307.03707 [hep-th]}}].

\bibitem{Belin:2020hea}
A.~Belin and J.~de~Boer, {{Random statistics of OPE coefficients and Euclidean
  wormholes}}, \href{http://dx.doi.org/10.1088/1361-6382/ac1082}{Class. Quant.
  Grav. {\bf 38}, 164001, 2021},
  [\href{http://arxiv.org/abs/arXiv:2006.05499}{{arXiv:2006.05499 [hep-th]}}].

\bibitem{Chandra:2022bqq}
J.~Chandra, S.~Collier, T.~Hartman and A.~Maloney, {{Semiclassical 3D gravity
  as an average of large-c CFTs}},
  \href{http://dx.doi.org/10.1007/JHEP12(2022)069}{JHEP {\bf 12}, 069, 2022},
  [\href{http://arxiv.org/abs/arXiv:2203.06511}{{arXiv:2203.06511 [hep-th]}}].

\bibitem{Collier:2019weq}
S.~Collier, A.~Maloney, H.~Maxfield and I.~Tsiares, {{Universal dynamics of
  heavy operators in CFT$_{2}$}},
  \href{http://dx.doi.org/10.1007/JHEP07(2020)074}{JHEP {\bf 07}, 074, 2020},
  [\href{http://arxiv.org/abs/arXiv:1912.00222}{{arXiv:1912.00222 [hep-th]}}].

\bibitem{Collier_2023}
S.~Collier, L.~Eberhardt and M.~Zhang, {Solving 3d gravity with virasoro tqft},
  \href{http://dx.doi.org/10.21468/scipostphys.15.4.151}{{\bf 15}, SciPost
  Physics, 2023}.

\bibitem{Collier:2024mgv}
S.~Collier, L.~Eberhardt and M.~Zhang, {{3d gravity from Virasoro TQFT:
  Holography, wormholes and knots}},  2024,
  [\href{http://arxiv.org/abs/arXiv:2401.13900}{{arXiv:2401.13900 [hep-th]}}].

\bibitem{Abajian:2023bqv}
J.~Abajian, F.~Aprile, R.~C. Myers and P.~Vieira, {{Correlation functions of
  huge operators in AdS$_{3}$/CFT$_{2}$: domes, doors and book pages}},
  \href{http://dx.doi.org/10.1007/JHEP03(2024)118}{JHEP {\bf 03}, 118, 2024},
  [\href{http://arxiv.org/abs/arXiv:2307.13188}{{arXiv:2307.13188 [hep-th]}}].

\bibitem{Chandra:2023dgq}
J.~Chandra and T.~Hartman, {{Toward random tensor networks and holographic
  codes in CFT}}, \href{http://dx.doi.org/10.1007/JHEP05(2023)109}{JHEP {\bf
  05}, 109, 2023},
  [\href{http://arxiv.org/abs/arXiv:2302.02446}{{arXiv:2302.02446 [hep-th]}}].

\bibitem{Chandra:2023rhx}
J.~Chandra, {{Euclidean wormholes for individual 2d CFTs}},
  \href{http://dx.doi.org/10.1007/JHEP04(2024)051}{JHEP {\bf 04}, 051, 2024},
  [\href{http://arxiv.org/abs/arXiv:2305.07183}{{arXiv:2305.07183 [hep-th]}}].

\bibitem{Yan:2023rjh}
C.~Yan, {{More on torus wormholes in 3d gravity}},
  \href{http://dx.doi.org/10.1007/JHEP11(2023)039}{JHEP {\bf 11}, 039, 2023},
  [\href{http://arxiv.org/abs/arXiv:2305.10494}{{arXiv:2305.10494 [hep-th]}}].

\bibitem{deBoer:2024kat}
J.~de~Boer, D.~Liska and B.~Post, {{Multiboundary wormholes and OPE
  statistics}},  2024,
  [\href{http://arxiv.org/abs/arXiv:2405.13111}{{arXiv:2405.13111 [hep-th]}}].

\bibitem{Belin:2023efa}
A.~Belin, J.~de~Boer, D.~L. Jafferis, P.~Nayak and J.~Sonner, {{Approximate
  CFTs and Random Tensor Models}},  2023,
  [\href{http://arxiv.org/abs/arXiv:2308.03829}{{arXiv:2308.03829 [hep-th]}}].

\bibitem{Jafferis:2024jkb}
D.~L. Jafferis, L.~Rozenberg and G.~Wong, {{3d Gravity as a random ensemble}},
  2024, [\href{http://arxiv.org/abs/arXiv:2407.02649}{{arXiv:2407.02649
  [hep-th]}}].

\bibitem{Saad:2019lba}
P.~Saad, S.~H. Shenker and D.~Stanford, {{JT gravity as a matrix integral}},
  2019, [\href{http://arxiv.org/abs/arXiv:1903.11115}{{arXiv:1903.11115
  [hep-th]}}].

\bibitem{Chandra:2022fwi}
J.~Chandra and T.~Hartman, {{Coarse graining pure states in AdS/CFT}},
  \href{http://dx.doi.org/10.1007/JHEP10(2023)030}{JHEP {\bf 10}, 030, 2023},
  [\href{http://arxiv.org/abs/arXiv:2206.03414}{{arXiv:2206.03414 [hep-th]}}].

\bibitem{Sasieta:2022ksu}
M.~Sasieta, {{Wormholes from heavy operator statistics in AdS/CFT}},
  \href{http://dx.doi.org/10.1007/JHEP03(2023)158}{JHEP {\bf 03}, 158, 2023},
  [\href{http://arxiv.org/abs/arXiv:2211.11794}{{arXiv:2211.11794 [hep-th]}}].

\bibitem{Balasubramanian:2022gmo}
V.~Balasubramanian, A.~Lawrence, J.~M. Magan and M.~Sasieta, {{Microscopic
  Origin of the Entropy of Black Holes in General Relativity}},
  \href{http://dx.doi.org/10.1103/PhysRevX.14.011024}{Phys. Rev. X {\bf 14},
  011024, 2024},
  [\href{http://arxiv.org/abs/arXiv:2212.02447}{{arXiv:2212.02447 [hep-th]}}].

\bibitem{Bah_2023}
I.~Bah, Y.~Chen and J.~Maldacena, {Estimating global charge violating
  amplitudes from wormholes},
  \href{http://dx.doi.org/10.1007/jhep04(2023)061}{{\bf 2023}, Journal of High
  Energy Physics, 2023}.

\bibitem{Chandra:2024vhm}
J.~Chandra, T.~Hartman and V.~Meruliya, {{Statistics of three-dimensional black
  holes from Liouville line defects}},  2024,
  [\href{http://arxiv.org/abs/arXiv:2404.15183}{{arXiv:2404.15183 [hep-th]}}].

\bibitem{Cotler_2021}
J.~Cotler and K.~Jensen, {Wormholes and black hole microstates in ads/cft},
  \href{http://dx.doi.org/10.1007/jhep09(2021)001}{{\bf 2021}, Journal of High
  Energy Physics, 2021}.

\bibitem{Marolf:2021kjc}
D.~Marolf and J.~E. Santos, {{AdS Euclidean wormholes}},
  \href{http://dx.doi.org/10.1088/1361-6382/ac2cb7}{Class. Quant. Grav. {\bf
  38}, 224002, 2021},
  [\href{http://arxiv.org/abs/arXiv:2101.08875}{{arXiv:2101.08875 [hep-th]}}].

\bibitem{Ghodsi:2022umc}
A.~Ghodsi, J.~K. Ghosh, E.~Kiritsis, F.~Nitti and V.~Nourry, {{Holographic QFTs
  on AdS$_{d}$, wormholes and holographic interfaces}},
  \href{http://dx.doi.org/10.1007/JHEP01(2023)121}{JHEP {\bf 01}, 121, 2023},
  [\href{http://arxiv.org/abs/arXiv:2209.12094}{{arXiv:2209.12094 [hep-th]}}].

\bibitem{Betzios:2019rds}
P.~Betzios, E.~Kiritsis and O.~Papadoulaki, {{Euclidean Wormholes and
  Holography}}, \href{http://dx.doi.org/10.1007/JHEP06(2019)042}{JHEP {\bf 06},
  042, 2019}, [\href{http://arxiv.org/abs/arXiv:1903.05658}{{arXiv:1903.05658
  [hep-th]}}].

\bibitem{Bak:2003jk}
D.~Bak, M.~Gutperle and S.~Hirano, {{A Dilatonic deformation of AdS(5) and its
  field theory dual}},
  \href{http://dx.doi.org/10.1088/1126-6708/2003/05/072}{JHEP {\bf 05}, 072,
  2003},
  [\href{http://arxiv.org/abs/arXiv:hep-th/0304129}{{arXiv:hep-th/0304129}}].

\bibitem{Bak:2007jm}
D.~Bak, M.~Gutperle and S.~Hirano, {{Three dimensional Janus and time-dependent
  black holes}}, \href{http://dx.doi.org/10.1088/1126-6708/2007/02/068}{JHEP
  {\bf 02}, 068, 2007},
  [\href{http://arxiv.org/abs/arXiv:hep-th/0701108}{{arXiv:hep-th/0701108}}].

\bibitem{Papadimitriou:2004rz}
I.~Papadimitriou and K.~Skenderis, {{Correlation functions in holographic RG
  flows}}, \href{http://dx.doi.org/10.1088/1126-6708/2004/10/075}{JHEP {\bf
  10}, 075, 2004},
  [\href{http://arxiv.org/abs/arXiv:hep-th/0407071}{{arXiv:hep-th/0407071}}].

\bibitem{Maldacena:1997re}
J.~M. Maldacena, {{The Large N limit of superconformal field theories and
  supergravity}}, \href{http://dx.doi.org/10.4310/ATMP.1998.v2.n2.a1}{Adv.
  Theor. Math. Phys. {\bf 2}, 231--252, 1998},
  [\href{http://arxiv.org/abs/arXiv:hep-th/9711200}{{arXiv:hep-th/9711200}}].

\bibitem{Miyaji:2015woj}
M.~Miyaji, T.~Numasawa, N.~Shiba, T.~Takayanagi and K.~Watanabe, {{Distance
  between Quantum States and Gauge-Gravity Duality}},
  \href{http://dx.doi.org/10.1103/PhysRevLett.115.261602}{Phys. Rev. Lett. {\bf
  115}, 261602, 2015},
  [\href{http://arxiv.org/abs/arXiv:1507.07555}{{arXiv:1507.07555 [hep-th]}}].

\bibitem{Balasubramanian:1999re}
V.~Balasubramanian and P.~Kraus, {{A Stress tensor for Anti-de Sitter
  gravity}}, \href{http://dx.doi.org/10.1007/s002200050764}{Commun. Math. Phys.
  {\bf 208}, 413--428, 1999},
  [\href{http://arxiv.org/abs/arXiv:hep-th/9902121}{{arXiv:hep-th/9902121}}].

\bibitem{Harlow:2011ny}
D.~Harlow, J.~Maltz and E.~Witten, {{Analytic Continuation of Liouville
  Theory}}, \href{http://dx.doi.org/10.1007/JHEP12(2011)071}{JHEP {\bf 12},
  071, 2011}, [\href{http://arxiv.org/abs/arXiv:1108.4417}{{arXiv:1108.4417
  [hep-th]}}].

\bibitem{Srednicki_1994}
M.~Srednicki, {Chaos and quantum thermalization},
  \href{http://dx.doi.org/10.1103/physreve.50.888}{Physical Review E {\bf 50},
  888–901, 1994}.

\bibitem{Cotler:2022rud}
J.~Cotler and K.~Jensen, {{A precision test of averaging in AdS/CFT}},  2022,
  [\href{http://arxiv.org/abs/arXiv:2205.12968}{{arXiv:2205.12968 [hep-th]}}].

\bibitem{Anous:2016kss}
T.~Anous, T.~Hartman, A.~Rovai and J.~Sonner, {{Black Hole Collapse in the 1/c
  Expansion}}, \href{http://dx.doi.org/10.1007/JHEP07(2016)123}{JHEP {\bf 07},
  123, 2016}, [\href{http://arxiv.org/abs/arXiv:1603.04856}{{arXiv:1603.04856
  [hep-th]}}].

\bibitem{Israel:1966rt}
W.~Israel, {{Singular hypersurfaces and thin shells in general relativity}},
  \href{http://dx.doi.org/10.1007/BF02710419}{Nuovo Cim. B {\bf 44S10}, 1,
  1966}. [Erratum: Nuovo Cim.B 48, 463 (1967)].

\bibitem{Affleck:1991tk}
I.~Affleck and A.~W.~W. Ludwig, {{Universal noninteger 'ground state
  degeneracy' in critical quantum systems}},
  \href{http://dx.doi.org/10.1103/PhysRevLett.67.161}{Phys. Rev. Lett. {\bf
  67}, 161--164, 1991}.

\bibitem{Cuomo:2021rkm}
G.~Cuomo, Z.~Komargodski and A.~Raviv-Moshe, {{Renormalization Group Flows on
  Line Defects}}, \href{http://dx.doi.org/10.1103/PhysRevLett.128.021603}{Phys.
  Rev. Lett. {\bf 128}, 021603, 2022},
  [\href{http://arxiv.org/abs/arXiv:2108.01117}{{arXiv:2108.01117 [hep-th]}}].

\bibitem{Hsin_2021}
P.-S. Hsin, L.~V. Iliesiu and Z.~Yang, {A violation of global symmetries from
  replica wormholes and the fate of black hole remnants},
  \href{http://dx.doi.org/10.1088/1361-6382/ac2134}{Classical and Quantum
  Gravity {\bf 38}, 194004, 2021}.

\bibitem{Chen_2021}
Y.~Chen and H.~W. Lin, {Signatures of global symmetry violation in relative
  entropies and replica wormholes},
  \href{http://dx.doi.org/10.1007/jhep03(2021)040}{{\bf 2021}, Journal of High
  Energy Physics, 2021}.

\bibitem{Gubser:1998bc}
S.~S. Gubser, I.~R. Klebanov and A.~M. Polyakov, {{Gauge theory correlators
  from noncritical string theory}},
  \href{http://dx.doi.org/10.1016/S0370-2693(98)00377-3}{Phys. Lett. B {\bf
  428}, 105--114, 1998},
  [\href{http://arxiv.org/abs/arXiv:hep-th/9802109}{{arXiv:hep-th/9802109}}].

\bibitem{Witten:1998qj}
E.~Witten, {{Anti-de Sitter space and holography}},
  \href{http://dx.doi.org/10.4310/ATMP.1998.v2.n2.a2}{Adv. Theor. Math. Phys.
  {\bf 2}, 253--291, 1998},
  [\href{http://arxiv.org/abs/arXiv:hep-th/9802150}{{arXiv:hep-th/9802150}}].

\bibitem{Freedman:2003ax}
D.~Z. Freedman, C.~Nunez, M.~Schnabl and K.~Skenderis, {{Fake supergravity and
  domain wall stability}},
  \href{http://dx.doi.org/10.1103/PhysRevD.69.104027}{Phys. Rev. D {\bf 69},
  104027, 2004},
  [\href{http://arxiv.org/abs/arXiv:hep-th/0312055}{{arXiv:hep-th/0312055}}].

\bibitem{Clark:2004sb}
A.~B. Clark, D.~Z. Freedman, A.~Karch and M.~Schnabl, {{Dual of the Janus
  solution: An interface conformal field theory}},
  \href{http://dx.doi.org/10.1103/PhysRevD.71.066003}{Phys. Rev. D {\bf 71},
  066003, 2005},
  [\href{http://arxiv.org/abs/arXiv:hep-th/0407073}{{arXiv:hep-th/0407073}}].

\bibitem{Maloney:2020nni}
A.~Maloney and E.~Witten, {{Averaging over Narain moduli space}},
  \href{http://dx.doi.org/10.1007/JHEP10(2020)187}{JHEP {\bf 10}, 187, 2020},
  [\href{http://arxiv.org/abs/arXiv:2006.04855}{{arXiv:2006.04855 [hep-th]}}].

\bibitem{Afkhami-Jeddi:2020ezh}
N.~Afkhami-Jeddi, H.~Cohn, T.~Hartman and A.~Tajdini, {{Free partition
  functions and an averaged holographic duality}},
  \href{http://dx.doi.org/10.1007/JHEP01(2021)130}{JHEP {\bf 01}, 130, 2021},
  [\href{http://arxiv.org/abs/arXiv:2006.04839}{{arXiv:2006.04839 [hep-th]}}].

\bibitem{Usatyuk:2024mzs}
M.~Usatyuk, Z.-Y. Wang and Y.~Zhao, {{Closed universes in two dimensional
  gravity}},  2024,
  [\href{http://arxiv.org/abs/arXiv:2402.00098}{{arXiv:2402.00098 [hep-th]}}].

\bibitem{Antonini:2023hdh}
S.~Antonini, M.~Sasieta and B.~Swingle, {{Cosmology from random entanglement}},
  \href{http://dx.doi.org/10.1007/JHEP11(2023)188}{JHEP {\bf 11}, 188, 2023},
  [\href{http://arxiv.org/abs/arXiv:2307.14416}{{arXiv:2307.14416 [hep-th]}}].

\bibitem{Benjamin:2023qsc}
N.~Benjamin, J.~Lee, H.~Ooguri and D.~Simmons-Duffin, {{Universal asymptotics
  for high energy CFT data}},
  \href{http://dx.doi.org/10.1007/JHEP03(2024)115}{JHEP {\bf 03}, 115, 2024},
  [\href{http://arxiv.org/abs/arXiv:2306.08031}{{arXiv:2306.08031 [hep-th]}}].

\end{thebibliography}\endgroup

\end{document}